\documentclass[%
 reprint,
 amsmath,amssymb,
 aps,
]{revtex4-2}

\usepackage{graphicx}
\usepackage{dcolumn}
\usepackage{bm}
\usepackage{color}
\usepackage{braket}
\usepackage{tikz}
\usetikzlibrary{arrows.meta}
\usepackage[percent]{overpic}
\usepackage{booktabs}
\usepackage{tabularx}
\usepackage[version=4]{mhchem}
\usepackage{fix-cm}
\usepackage{mwe}
\newcommand{\panelLabel}[1]{%
  \bfseries\fontsize{10}{14}\selectfont #1%
}

\makeatother

\begin{document}

\preprint{APS/123-QED}

\title{A boson representation of the electronic structure calculation based on the quantum Krylov subspace algorithm}

\author{Hiroki Katow}
\thanks{Two authors contributed equally to this work.}
\email{hiroki.kato@aist.go.jp}
\author{Hisao Nakamura}%
\thanks{Two authors contributed equally to this work.}
\email{hs-nakamura@aist.go.jp}
\affiliation{%
Material DX Research Center, Advanced Institute of Industrial Science and Technology (AIST), Tsukuba, Japan 
}%

\date{\today}

\begin{abstract}
Quantum calculations are expected to broaden the range of strongly correlated phenomena in solids accessible to first-principles calculations.
The quantum Krylov diagonalization method, or in a broader sense, the quantum subspace expansion method (QSE) offers an advantage over other variational methods in terms of numerical stability.
In this work, we propose a boson representation of electronic systems with the aim of reducing QSE quantum circuit depth, 
where electron-hole pair bosons serve as the fundamental degrees of freedom. 
We benchmark the boson representation in comparison with a corresponding fermionic QSE and CCSD for an equidistant and a dimerized configuration of a periodic hydrogen chain \ce{(H2)_n}. 
Despite the drastic approximation we reveal the boson QSE scheme provides practically useful accuracy in the ground state energy.
Our results support its applicability to realistic quasi-two- and three-dimensional electronic systems.
\end{abstract}

\maketitle


\section{Introduction}
First-principles electronic structure calculations play a central role in materials discovery through the prediction and screening of candidate compounds. 
Although methods based on density functional theory and related approximations enable simulations of solids at practical computational cost, 
the accurate treatment of strong electron correlation remains a fundamental challenge.

Quantum computing offers a promising route toward highly accurate electronic structure calculations by spanning a huge Hilbert space on quantum hardware. 
Rapid advances in quantum hardware have led to increasing system sizes and improved gate fidelities and coherence times. 
However, quantum error correction on a practical scale has yet to be realized. 
Consequently, substantial effort has been devoted to the development of resource-efficient and numerically robust quantum algorithms 
suitable for noisy intermediate-scale quantum (NISQ) devices and, ultimately, early fault-tolerant quantum computers (FTQC).

The present study aims at developing a numerically efficient {\it ab initio} model for the electronic structure calculations of solid-state systems. 
Considerable efforts have been devoted to reducing the computational cost of quantum algorithms; 
however, we anticipate sampling more than thousands of orbitals required for the convergence of physical observables in realistic solids will remain challenging even in the early FTQC era.
In classical algorithms, computational efficiency is often achieved by systematically incorporating a limited class of correlation effects in its framework. 
A natural question arises as to whether a similar strategy can be exploited in quantum algorithms. 
One example is the DFT+GW+BSE framework, which is one of the most successful approaches for predicting the optical properties of band insulators.
Within the density functional theory (DFT) \cite{Hohenberg_1964,Kohn_1965}, Kohn-Sham orbitals provide a mean-field description of the electronic ground state. 
Building on the DFT solution, the GW approximation \cite{Hedin_1965,Hybertsen_1986} incorporates screening-induced self-energy corrections and yields improved quasiparticle spectra, 
while the Bethe-Salpeter equation (BSE) \cite{Rohlfing_2000} accounts for electron-hole correlations.
It results in the excitonic effects that govern the optical gaps.
Despite its capability of quantitative prediction, the DFT+GW+BSE framework remains computationally demanding and the type of correlation is limited. 
It therefore cannot describe higher-order correlations such as biexcitonic effects. 
Nevertheless this framework provides an insight that the electron-hole pair serves as a principal carrier of many-body correlation for band insulators.  

Motivated by this observation, we adopt electron-hole pairs as the effective bosonic degrees of freedom.
We transform the original fermionic many-electron Hamiltonian into an effective bosonic Hamiltonian, which enables us to implement the quantum subspace expansion method (QSE) \cite{Parrish_2019,Huggins_2020,Seki_2021} using a shallower quantum circuit.
Within this framework, a quantum computer is employed to generate an efficient, generally nonorthogonal, subspace basis for a quantum many-body system and to evaluate the matrix elements of the Hamiltonian and overlap (Gram) matrices. 
The resulting generalized eigenvalue problem is subsequently solved on a classical computer.
Numerous derivatives differing in the construction of the subspace basis have been developed, 
including qEOM\cite{Ollitrault_2020}, Chebyshev quantum Krylov\cite{Kirby_2023}, Gaussian-power quantum Krylov\cite{Zhang_2024}, quantum filter diagonalization\cite{Parrish_2019}, 
multireference selected quantum Krylov\cite{Stair_2020}, quantum imaginary-time evolution\cite{Motta_2019}, and quantum selected configuration interaction\cite{Mikkelsen_2025,Kanno_2026}. 
Comprehensive overviews of these developments can be found in a recent review article\cite{Motta_2024}. 
A recent implementation of the QSE on real quantum devices can be found in ref.~\cite{Yoshioka_2025}.
We employ the quantum Krylov diagonalization (QKD) scheme \cite{Cortes_2022} where a nonorthogonal basis set is generated by applying the bosonic time-evolution operator to a reference state.
A quantum computer returns Hamiltonian and overlap matrices which constitute a generalized eigenvalue problem.
A key advantage of the boson representation in this scheme is that it is mapped onto a spin system in one-to-one correspondence preserving the locality of Hamiltonian. 
This contrasts with fermionic encodings, where the nonlocal fermion-to-spin mapping requires strings of CNOT gates whose length may scale with system size \cite{Whitfield_2011}. 
The bosonic formulation therefore offers the prospect of substantially reducing quantum-circuit depth.


The QSE method should be contrasted with the variational quantum algorithms (VQAs).
VQAs constitutes early attempts for computing eigenstates of quantum many-body systems on noisy intermediate-scale quantum (NISQ) devices. 
In VQAs, a trial wave function is represented by a parameterized quantum circuit.
The energy is evaluated on a quantum processor while the circuit parameters are optimized classically. 
Since their introduction by Peruzzo et al.~\cite{Peruzzo_2014} and subsequent development by McClean et al.~\cite{McClean_2016}, 
VQAs have been widely applied to quantum chemistry and many-body problems across a broad range of physical systems \cite{Mitarai_2019, Liu_2020,Yoshioka_2022,Manrique_2020, Yoshida_2024,Yoshida_2026}. 
A comprehensive review is available in ref.~\cite{Tilly_2022}.
A major limitation of VQAs is the barren plateau problem (BP), in which cost-function gradients vanish exponentially with system size, making optimization increasingly challenging. 
Developing strategies to mitigate BP remains an active area of research \cite{Larocca_2025}.
The QSE methods, on the other hand, do not involve nonlinear optimization of variational circuit parameters. 
The QSE thus has a prominent feature of being free from BP and tends to exhibit robustness against hardware noise. 

The construction of this paper is as follows.
In Sec.~\ref{sec:boson} we introduce a concept of the electron-hole pair boson representation of the electronic many-body Hamiltonian.
An additional approximation discarding the spin degrees of freedom is introduced in order to reduce the number of qubits in Sec.~\ref{subsec:spinless_boson}.
A boson-spin mapping is given by the Holstein-Primakoff transformation in Sec.~\ref{subsec:boson_spin}.
We show a Pauli matrix expression of electron many-body Hamiltonian for selected terms in Sec.~\ref{subsec:boson_Hamiltonian}.
A quantum circuit expression of a bosonic operator is given in Sec.~\ref{subsec:boson_circuit}.  
The QKD method and our measurement protocol for matrix elements are briefly described in Sec.~\ref{sec:QSE}.
We apply the spinless boson QSE formalism to ground-state calculations for the periodic hydrogen chains in Sec.~\ref{sec:results}. 
Neglecting spin degrees of freedom throughout, we compare the resulting spinless boson QSE with its fermionic counterpart.
Despite the rather drastic approximation its ground state energy shows fairly good agreement with CCSD, although a possible cancellation of competing errors should be taken into account.
These results support its applicability to realistic quasi-two- and three-dimensional electronic systems.
Possibility of further improvements of the boson representation is discussed through Sec.~\ref{sec:discussion}.
Concluding remarks are presented in Sec.~\ref{sec:concluding_remarks}.

\section{Boson representation of electronic state}
\label{sec:boson}
The main concept of this work is a boson representation which leads to a great reduction of the circuit depth in QSE implementation.
We introduce a mapping of an electron-hole pair excitation (deexcitation) operator onto a boson operator as:   
\begin{eqnarray}
    a_{\mathbf{p}c\sigma}^\dagger a_{\mathbf{k}v\rho} \to \beta^\dagger_{\mathbf{p}c\sigma\mathbf{k}v\rho} ,\;
    a_{\mathbf{k}v\rho}^\dagger a_{\mathbf{p}c\sigma} \to \beta_{\mathbf{p}c\sigma\mathbf{k}v\rho}.
    \label{eq:def_beta}
\end{eqnarray}
A pair of indices $(\mathbf{k},v,\rho)$ ($(\mathbf{p},c,\sigma)$) indicates the crystal momentum, band index, and spin of the hole (electron), respectively.
We place the boson Eq.~(\ref{eq:def_beta}) as a building block of the rest of operators.
For this purpose we introduce another approximation for valence-valence excitation and conduction-conduction excitation operators as: 
\begin{eqnarray}
    a_{\mathbf{k}_1v_1\rho_1}^\dagger a_{\mathbf{k}_2v_2\rho_2} 
    &\sim & \frac{1}{n_k n_c n_s}\sum_{\mathbf{p}c \sigma} a^\dagger_{\mathbf{k}_1v_1\rho_1} a_{\mathbf{p}c \sigma} a^\dagger_{\mathbf{p}c\sigma} a_{\mathbf{k}_2v_2\rho_2} \nonumber\\
     &\to& \frac{1}{n_k n_cn_s} \sum_{\mathbf{p}c\sigma} \beta_{\mathbf{p}c\sigma\mathbf{k}_1v_1\rho_1} \beta^\dagger_{\mathbf{p}c\sigma\mathbf{k}_2v_2\rho_2}
     \label{eq:def_vv}
\end{eqnarray}
and
\begin{eqnarray}
    a_{\mathbf{p}_1c_1\sigma_1}^\dagger a_{\mathbf{p}_2c_2\sigma_2} 
    &\sim & \frac{1}{n_k n_vn_s}\sum_{\mathbf{k}v\rho}  a_{\mathbf{p}_1c_1 \sigma_1}^\dagger a_{\mathbf{k}v\rho} a^\dagger_{\mathbf{k}v\rho} a_{\mathbf{p}_2c_2 \sigma_2} \nonumber \\
     &\to & \frac{1}{n_k n_vn_s} \sum_{\mathbf{k}v\rho} \beta^\dagger_{\mathbf{p}_1c_1\sigma_1\mathbf{k}v\rho} \beta_{\mathbf{p}_2c_2\sigma_2\mathbf{k}v\rho}.
      \label{eq:def_cc}
\end{eqnarray}
This treatment is based on an assumption that the transition processes expressed by Eqs.~(\ref{eq:def_vv}) and (\ref{eq:def_cc}) are mediated by electron-hole pair excitations.   
\subsection{Boson vacuum and commutation relation}
\label{subsec:vac_CR}
In the following we examine the validity of our boson representation.
We take the Hartree-Fock ground state as a reference state to which we apply the time evolution operator.
Let such a state be denoted by $\ket{0}$.
We suppose that the reference state is gapped and all valence bands are fully occupied. 
Then the boson creation and annihilation operators Eq.~(\ref{eq:def_beta}) act on $\ket{0}$ as
\begin{eqnarray}
    \bra{0}\beta^\dagger_{\mathbf{p}c\sigma\mathbf{k}v\rho} = 0\\
    \beta_{\mathbf{p}c\sigma\mathbf{k}v\rho} \ket{0} = 0.
    \label{eq:vacuum}
\end{eqnarray} 
The reference state $\ket{0}$ thus becomes a boson vacuum in 
this representation.
The commutation relation between electron-hole pair excitation operators is
\begin{eqnarray}
    &&[a^\dagger_{\mathbf{k}_1v_1\rho_1}a_{\mathbf{p}_1c_1\sigma_1},a^\dagger_{\mathbf{p}_2c_2\sigma_2}a_{\mathbf{k}_2v_2\rho_2}]\nonumber \\
    &&=\delta_{\mathbf{p}_1\mathbf{p}_2}\delta_{c_1c_2}\delta_{\sigma_1\sigma_2}\delta_{\mathbf{k}_1\mathbf{k}_2}\delta_{v_1v_2}\delta_{\rho_1\rho_2}\nonumber \\
    &&+\delta_{\mathbf{p}_1\mathbf{p}_2}\delta_{c_1c_2}\delta_{\sigma_1\sigma_2}a_{\mathbf{k}_2v_2\rho_2}a_{\mathbf{k}_1v_1\rho_1}^\dagger \nonumber\\
    &&-\delta_{\mathbf{k}_1\mathbf{k}_2}\delta_{v_1v_2}\delta_{\rho_1\rho_2}a^\dagger_{\mathbf{p}_2c_2\sigma_2}a_{\mathbf{p}_1c_1\sigma_1}.
    \label{eq:fermion_CR}
\end{eqnarray}
Because of the last two terms the operators in Eq.~(\ref{eq:def_beta}) cannot be exact bosonic operators in general cases.
Our boson picture is justified when the expectation values of the two terms are negligible, namely $|\braket{a_{\mathbf{k}_1v_1\rho_1} a^\dagger_{\mathbf{k}_2v_2\rho_2}}| \ll 1 $ and 
$|\braket{a^\dagger_{\mathbf{p}_2c_2\sigma_2}a_{\mathbf{p}_1c_1\sigma_1}}|\ll 1$, which is a straightforward expression of our assumption that the electron-hole pair excitation density remains perturbative.
As long as this condition is fulfilled the electron-hole pair boson Eq.~(\ref{eq:def_beta}) satisfies a complete boson commutation relation:
\begin{eqnarray}
    [\beta_{\mathbf{p}_1c_1\sigma_1\mathbf{k}_1v_1},\beta^\dagger_{\mathbf{p}_2c_2\sigma_2\mathbf{k}_2v_2}]
    =\delta_{\mathbf{p}_1\mathbf{p}_2}\delta_{c_1c_2}\delta_{\sigma_1\sigma_2}\delta_{\mathbf{k}_1\mathbf{k}_2}\delta_{v_1v_2}\delta_{\rho_1\rho_2}
    \label{eq:boson_CR}.\nonumber \\
\end{eqnarray}
\subsection{Spinless boson representation}
\label{subsec:spinless_boson}
According to the definition Eq.~(\ref{eq:def_beta}) the number of boson modes $N_B$ is given by $N_B=n_s^2n_k^2n_vn_c$.
The rapid growth of $N_B$ make it hard to run QSE on a quantum circuit simulator.
Here we carefully omit the spin degrees of freedom in order to sample multiple $k$ points and bands.
We first ignore the spin triplet states. 
This is reasonable if the spin-orbit coupling is negligible.
We then introduce a spinless boson operator as follows:
\begin{eqnarray}
    \beta_{\mathbf{p}c\mathbf{k}v}\equiv\frac{1}{2}(\beta_{\mathbf{p}c\uparrow\mathbf{k}v\uparrow}+\beta_{\mathbf{p}c\downarrow\mathbf{k}v\downarrow}).
    \label{eq:def_spinless_boson}
\end{eqnarray}
This approximation would be plausible if no magnetic order is expected.
Nonetheless Eq.~(\ref{eq:def_spinless_boson}) may be a good approximation for the case that 
the long-range dielectric screening by singlet electron-hole pair excitation dominates the total energy correction like the GW approximation.
We note that the spinless boson representation Eq.~(\ref{eq:def_spinless_boson}) is introduced to reduce the number of qubits and  independent of the bosonization scheme itself.
With sufficient computational resources we can discard this restriction so that the boson operators describe either singlet or triplet excitations as originally formulated in Eq.~(\ref{eq:def_beta}).

\subsection{Spin-boson mapping}
\label{subsec:boson_spin}
 We provide a mapping of the electron-hole pair boson Eq.~(\ref{eq:def_beta}) onto Pauli operators through 
the Holstein-Primakoff transformation \cite{Holstein_1940} :
\begin{eqnarray}
    S_{+}&\equiv&\sqrt{2S}\cdot \beta^\dagger\left(1-\frac{1}{2S}\beta^\dagger \beta\right)^{1/2}\\
    S_{-}&\equiv&\sqrt{2S}\left(1-\frac{1}{2S}\beta^\dagger \beta\right)^{1/2}\cdot \beta\\
    S_{z}&\equiv&-S+\beta^\dagger \beta.
    \label{eq:def_HP}
\end{eqnarray}
for maximum angular momentum $S=1/2$.
The double occupation of the same boson mode $(\beta^\dagger)^2\ket{0}$ is therefore vanishing. 
The square-root factors involving boson operators in Eq.~(\ref{eq:def_HP}) can be evaluated by Taylor expansion.
We truncate the expansion at the lowest order. 
The boson-Pauli operator one-to-one correspondence is then given by
\begin{eqnarray}
    \beta^\dagger &\to& \sigma_+ \\
    \beta &\to& \sigma_-,\\
    \beta^\dagger\beta-1/2 &\to& \sigma_z.
    \label{eq:def_boson_to_spin}
\end{eqnarray}
This contrasts to the fermionic system where the Fermion-Pauli operator mappings, e.g. Jordan-Wigner transformation \cite{Jordan_1928}  are always accompanied by a product of non-local Pauli operators.
We will see in the next subsection that the absence of non-local operators plays a central role in substantially reducing the circuit depth of the Coulomb interaction term. 
Given the number of $k$ point sampling $n_k$ and valence (conduction) band $n_v$ ($n_c$), 
the number of electron-hole pair boson is $n_k^2n_vn_c$. 
Accordingly, $n_k^2n_vn_c$ qubits are required to express the whole electronic system.

\subsection{Hamiltonian of the boson approximation}
\label{subsec:boson_Hamiltonian}
Here we derive the Hamiltonian of boson approximation.
As we stated above, the reference state $\ket{0}$ is taken to be a Hartree-Fock ground state.
We expect that the deviation of exact ground state from the reference state is perturbative.
We therefore split the total Hamiltonian $\hat{H}$ as
\begin{eqnarray}
    \hat{H} = \hat{H}_{\textrm{HF}}+\Delta\hat{H}
    \label{eq:def_Hamiltonian}
\end{eqnarray}
where
\begin{eqnarray}
    \Delta\hat{H} \equiv \hat{H} - \hat{H}_{\textrm{HF}}.
    \label{eq:def_dH}
\end{eqnarray}
$\hat{H}_{\textrm{HF}}$ is the Hartree-Fock Hamiltonian.
In this expression the kinetic energy, electron-ion interaction, and ion-ion interaction 
are all included in the one-particle orbital energy $\epsilon^{\rm{HF}}_{\mathbf{k}m\sigma}$ given by Hartree-Fock calculation.
Since we exclude the spin degrees of freedom as we see in Eq.~(\ref{eq:def_spinless_boson}), the weight associated with spin degeneracy $n_s$ has to be appropriately multiplied.
Transforming each term of Eq.~(\ref{eq:def_dH}) into bosonic operator is a tedious and complex procedure.  
In this subsection we only provide several examples to see the results of bosonization.
The complete Hamiltonian is given in Appendix \ref{sec:hamiltonian}.

Mapping of the one-particle orbital energy term is the simplest example.
Using Eq.~(\ref{eq:def_vv}), we have
\begin{eqnarray}
    &&\sum_{\mathbf{k},v,\rho}\epsilon^{\rm{HF}}_{\mathbf{k}v\rho}c^\dagger_{\mathbf{k}v\rho}c_{\mathbf{k}v\rho}\sim n_s\sum_{\mathbf{k},v}\epsilon^{\rm{HF}}_{\mathbf{k}v}c^\dagger_{\mathbf{k}v}c_{\mathbf{k}v}\nonumber\\
    &&\to \frac{n_s}{n_kn_c}\sum_{\mathbf{k},v,\mathbf{p},c}\epsilon^{\rm{HF}}_{\mathbf{k}v}\beta_{\mathbf{p}c\mathbf{k}v}\beta_{\mathbf{p}c\mathbf{k}v}^\dagger \nonumber\\
&&\to \frac{n_s}{n_kn_c}\sum_{\mathbf{k},v,\mathbf{p},c}\epsilon^{\rm{HF}}_{\mathbf{k}v}\sigma^-_{\mathbf{Q}}\sigma^+_{\mathbf{Q}}=\frac{n_s}{n_kn_c}\sum_{\mathbf{Q}}\epsilon_{\mathbf{k}v}^{\rm{HF}}\begin{bmatrix}0 & 0 \\ 0 & 1\end{bmatrix}_{\mathbf{Q}}.
\label{eq:def_H_ekm}
\end{eqnarray}
$\epsilon^{\rm{HF}}_{\mathbf{k}v}$ is the one-particle orbital energy given by Hartree-Fock calculation.
$\mathbf{Q}=(\mathbf{k},v,\mathbf{p},c)$ is a collective index uniquely specifying a corresponding qubit.
Since we only consider spin-singlet pairs, a factor $n_s$ in the denominator of Eq.~(\ref{eq:def_vv}) is absent.
The spin weight $n_s$ is multiplied based on an assumption that the state $(\mathbf{k},v)$ is doubly occupied.
A diagonal contribution is always accompanied by a factor $n_s$. 
The off-diagonal contribution of one-body Hamiltonian $h_1$ is transformed in a different manner:
\begin{eqnarray}
    h_{1;\mathbf{k}v\sigma\mathbf{p}c\sigma}c_{\mathbf{k}v\sigma}^\dagger c_{\mathbf{p}c\sigma}&\sim& \hat{h}_{1;\mathbf{k}v\mathbf{p}c}c_{\mathbf{k}v}^\dagger c_{\mathbf{p}c} \nonumber \\
    &\to& h_{1;\mathbf{k}v\mathbf{p}c}\beta_{\mathbf{k}v\mathbf{p}c} \nonumber \\
    &\to& h_{1;\mathbf{Q}}\sigma^-_{\mathbf{Q}}.
\end{eqnarray}
Here the spin weight $n_s$ is not multiplied. 

Lastly we show a double excitation term of two-body Hamiltonian $h_2$. A double excitation from a valence band to a conduction band should be like
\begin{eqnarray}
    &&h_{2;\mathbf{k}v\mathbf{p}c}c^\dagger_{\mathbf{p}c\uparrow}c_{\mathbf{k}v\uparrow}c^\dagger_{\mathbf{p}c\downarrow}c_{\mathbf{k}v\downarrow}\nonumber \\
    &&= \frac{1}{2} h_{2;\mathbf{k}v\mathbf{p}c} (c^\dagger_{\mathbf{p}c\uparrow}c_{\mathbf{k}v\uparrow}+c^\dagger_{\mathbf{p}c\downarrow}c_{\mathbf{k}v\downarrow})^2\nonumber \\
    &&\to 2 h_{2;\mathbf{k}v\mathbf{p}c}\beta^\dagger_{\mathbf{p}c\mathbf{k}v}\nonumber \\
    &&\to 2h_{2;\mathbf{Q}}\sigma^+_{\mathbf{Q}}.
\end{eqnarray}
Although the expression of boson Hamiltonian is not unique and its derivation is complicated, one can derive it following the mapping above.
\subsection{Quantum circuit of boson representation}
\label{subsec:boson_circuit}

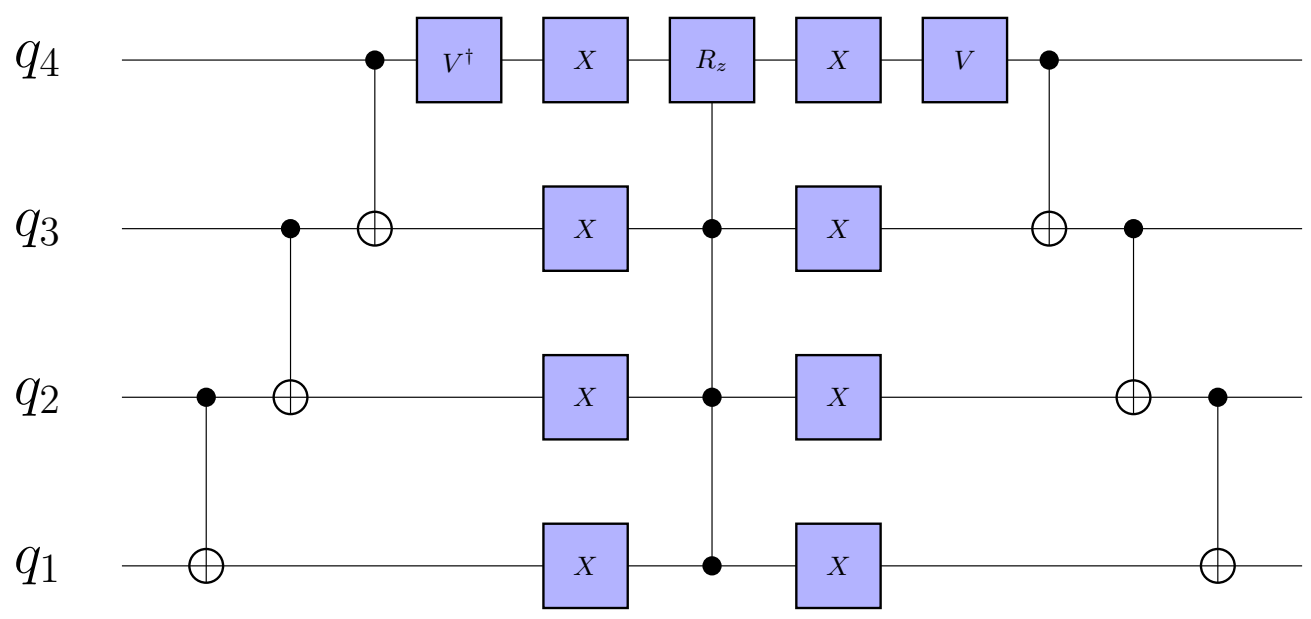
\begin{figure}[tb]
    \centering
    \resizebox{\linewidth}{!}{    
	\begin{tikzpicture}
    \node[font=\fontsize{20}{24}\selectfont] at (-1,-2) {$q_1$};
    \node[font=\fontsize{20}{24}\selectfont] at (-1,0) {$q_2$};
    \node[font=\fontsize{20}{24}\selectfont] at (-1,2) {$q_3$};
    \node[font=\fontsize{20}{24}\selectfont] at (-1,4) {$q_4$};
    \draw (0, -2) -- (14,-2);
    \draw (0, 0) -- (14,0);
    \draw (0, 2) -- (14,2);
    \draw (0, 4) -- (14,4);

    \filldraw[fill opacity = 0.0, draw=black, thick] (1,-2) circle (0.2);
    \filldraw[fill=black, draw=black, thick] (1,0) circle (0.1);
    \draw (1.0,-2.2) -- (1,0.0) ;
    \filldraw[fill opacity = 0.0, draw=black, thick] (2,0) circle (0.2);
    \filldraw[fill=black, draw=black, thick] (2,2) circle (0.1);
    \draw (2.0,-0.2) -- (2,2.0) ;
    \filldraw[fill opacity = 0.0, draw=black, thick] (3,2) circle (0.2);
    \filldraw[fill=black, draw=black, thick] (3,4) circle (0.1);
    \draw (3.0,1.8) -- (3,4.0) ;

    \filldraw[fill=blue!30, draw=black, thick] (3.5,3.5) rectangle (4.5,4.5);
    \node at (4.0,4.0) {$V^\dagger$};

    \filldraw[fill=blue!30, draw=black, thick] (5.0,3.5) rectangle (6.0,4.5);
    \node at (5.5,4.0) {$X$};
    \filldraw[fill=blue!30, draw=black, thick] (5.0,1.5) rectangle (6.0,2.5);
    \node at (5.5,2.0) {$X$};
    \filldraw[fill=blue!30, draw=black, thick] (5.0,-0.5) rectangle (6.0,0.5);
    \node at (5.5,0.0) {$X$};
    \filldraw[fill=blue!30, draw=black, thick] (5.0,-2.5) rectangle (6.0,-1.5);
    \node at (5.5,-2.0) {$X$};

    \filldraw[fill=black, draw=black, thick] (7,-2) circle (0.1);
    \filldraw[fill=black, draw=black, thick] (7,0) circle (0.1);
    \filldraw[fill=black, draw=black, thick] (7,2) circle (0.1);
    \draw (7.0,-2.1) -- (7,4) ;
    \filldraw[fill=blue!30, draw=black, thick] (6.5,3.5) rectangle (7.5,4.5);
    \node at (7.0,4.0) {$R_z$};

    \filldraw[fill=blue!30, draw=black, thick] (8.0,3.5) rectangle (9.0,4.5);
    \node at (8.5,4.0) {$X$};
    \filldraw[fill=blue!30, draw=black, thick] (8.0,1.5) rectangle (9.0,2.5);
    \node at (8.5,2.0) {$X$};
    \filldraw[fill=blue!30, draw=black, thick] (8.0,-0.5) rectangle (9.0,0.5);
    \node at (8.5,0.0) {$X$};
    \filldraw[fill=blue!30, draw=black, thick] (8.0,-2.5) rectangle (9.0,-1.5);
    \node at (8.5,-2.0) {$X$};

    \filldraw[fill=blue!30, draw=black, thick] (9.5,3.5) rectangle (10.5,4.5);
    \node at (10.0,4.0) {$V$};

    \filldraw[fill opacity = 0.0, draw=black, thick] (11,2) circle (0.2);
    \filldraw[fill=black, draw=black, thick] (11,4) circle (0.1);
    \draw (11.0,1.8) -- (11,4.0) ;
    \filldraw[fill opacity = 0.0, draw=black, thick] (12,0) circle (0.2);
    \filldraw[fill=black, draw=black, thick] (12,2) circle (0.1);
    \draw (12.0,-0.2) -- (12,2.0) ;
    \filldraw[fill opacity = 0.0, draw=black, thick] (13,-2) circle (0.2);
    \filldraw[fill=black, draw=black, thick] (13,0) circle (0.1);
    \draw (13.0,-2.2) -- (13,0.0) ;
    
	\end{tikzpicture}}
    \caption{A quantum circuit expression of $U_4$ in Eq.~(\ref{eq:def_U4}) which corresponds to a two-body Coulomb interaction term involving four qubits $q_1, q_2,\cdots, q_4$.}
    \label{fig:fig_U4}
\end{figure}

In the previous subsection, we provided several examples of bosonic representation of Eq.~(\ref{eq:def_dH}).
The aim of this subsection is to provide a circuit representation of the time-evolution operator of the bosonic Hamiltonian. 
Here we provide an example of a term $\hat{h}_4$ which incorporates a four-qubit operation.
It comes when four orbital indices of the electron repulsion integral (ERI) all belong to valence (conduction) bands.
By appropriately converting $\sigma^+$ and $\sigma^-$ using $X$ gates, $\hat{h}_4$ has a form
\begin{eqnarray}
    \hat{h}_4 = h_4 \sigma_1^+ \sigma_2^+\sigma_3^+\sigma_4^+ + h_4^* \sigma_1^- \sigma_2^-\sigma_3^-\sigma_4^-.
    \label{eq:def_h4}
\end{eqnarray}
Here $h_4\in \mathbb{C}$ and $h_4^*$ is the complex conjugate.
Then its time-evolution operator $\hat{U}_4(t)$ is
\begin{eqnarray}
    \hat{U}_4(t) &\equiv& e^{-it\hat{h}_4}\nonumber\\
    &=&\land_1(X_2)\land_2(X_3)\land_3(X_4)\nonumber\\
    &&\times X_1X_2X_3X_4\nonumber\\
    &&\times V_4^\dagger(\theta) \land_{123}(R_{z;4}(2t\rho)) V_4(\theta)\nonumber \\
    &&\times X_1X_2X_3X_4\nonumber\\
    &&\times \land_3(X_4)\land_2(X_3)\land_1(X_2).
    \label{eq:def_U4}
\end{eqnarray}
$\land_i(X_j)$ is the CNOT gate. $\land_{123}(R_{z;4})$ is the multicontrolled-$R_z$ gate where qubits $q_1$, $q_2$, and $q_3$ are control-qubits and qubit $q_4$ is the target qubit. 
The gate operation $V(\theta)$ is defined by
\begin{eqnarray}
    V(\theta)\equiv \frac{1}{\sqrt{2}}\begin{bmatrix}1 & 1 \\ e^{i\theta} & -e^{i\theta}\end{bmatrix}=G(\theta/2)R_z(\theta)H,
\end{eqnarray}
where $G$ is the phase gate and $H$ is the Hadamard gate.
$\theta$ is the argument of $h_4$ and $\rho$ is its amplitude, namely $h_4=|\rho|e^{i\theta}$.
$V(\theta)$ satisfies the following relation:
\begin{eqnarray}
    h_4\sigma_++h_4^*\sigma_-=
    \begin{bmatrix}
        0 & h_4^*\\
        h_4 & 0 
    \end{bmatrix}
    = V(\theta)
    \begin{bmatrix}
        \rho & 0 \\
        0 & -\rho
    \end{bmatrix}
    V^\dagger(\theta).
\end{eqnarray}
The corresponding quantum circuit is depicted in Fig.~\ref{fig:fig_U4}. 
Now we highlight that $U_4$ only incorporates four boson modes.
In the original fermionic representation, the equivalent term involves $O(n_kn_b)$ qubits where $n_b=n_c+n_v$ at maximum due to the Pauli string assuring the anticommuting relation.
Accordingly an equivalent circuit has $O(n_kn_b)$ depth in fermionic representation, which is now reduced to $O(1)$ circuit depth in our boson representation.  

\section{Quantum Krylov Subspace Algorithm}
\label{sec:QSE}
\subsection{Quantum Krylov subspace algorithm}
\label{subsec:QSE}
A standard formulation of the QSE method constructs a generalized eigenvalue problem 
\begin{eqnarray}
    \bm{H}\bm{C}=E\bm{S}\bm{C}
\end{eqnarray} 
to obtain the ground-state energy of the system.
$\bm{H}$ is the system Hamiltonian, $\bm{C}$ is the eigenvector in a subspace spanned by a nonorthogonal basis set
 whose overlap matrix is $\bm{S}$.
In the quantum Krylov subspace algorithm the nonorthogonal basis set $\{\ket{t_1},\ket{t_2},\cdots\}$ is composed of state vectors $\ket{t}$ at time $t$ evolved using the time-evolution operator as:
\begin{eqnarray}
    \ket{t}=e^{-i\hat{H}t}\ket{0}.
\end{eqnarray}
Here $\ket{0}$ is the initial state of system.
The Hamiltonian $\bm{H}$ and overlap matrix $\bm{S}$ are then 
\begin{eqnarray}
    H_{ij} = \bra{t_i}\hat{H}\ket{t_j}
    \label{eq:def_Hij}
\end{eqnarray}
and
\begin{eqnarray}
    S_{ij} = \langle t_i|t_j\rangle.
    \label{eq:def_Sij}
\end{eqnarray}


There is an alternative construction which diagonalizes a matrix-valued function of Hamiltonian $f(\hat{H})$ instead of $\hat{H}$ itself.
As long as the inverse of function $f(E)$ is well-defined, we can recover the eigenvalue $E$ from $f(E)$.
In this study we choose such $f(\hat{H})$ to be
\begin{eqnarray}
    f(\tau;\hat{H})=e^{-i\tau\hat{H}},
    \label{eq:def_f}
\end{eqnarray}
where $\tau$ is real.
This choice is referred as the KDM-U method \cite{Cortes_2022}.
In this work we sample $t_i$ from a uniform time grid for simplicity.
We note this sampling method is not a unique choice.
It has been reported that a non-uniform sampling can improve the convergence of energy with respect to the number of basis states \cite{Jamet_2022}.
Equations (\ref{eq:def_Hij}) and (\ref{eq:def_Sij}), however, have a Toeplitz structure which greatly reduces the cost to prepare the matrix elements. 
The overlap matrix Eq.~(\ref{eq:def_Sij}) can easily be overcomplete when the number of time-grid points is increased.
We apply canonical orthogonalization to $\bm{S}$ to exclude the redundancy of nonorthogonal basis set and achieve numerical stability.
The time evolution operator $e^{-i\hat{H}t}$ is implemented on a quantum circuit by the second-order Suzuki-Trotter decomposition.
Writing the second-order Suzuki-Trotter decomposition as $D_2(t) = (e^{-it\hat{h}_M/2m}\cdots e^{-it\hat{h}_{1}/m}  \cdots e^{-it\hat{h}_M/2m})^m$ 
for $\hat{H}=\sum_i^M\hat{h}_i$, we implement
\begin{eqnarray}
    S_{ij}&=&\bra{0} D_2(t_i-t_j)\ket{0}\\
    F_{ij}&=&\bra{0} D_2(\tau+t_i-t_j)\ket{0}.
\end{eqnarray}
Now we have several parameters whose convergence needs to be checked: {\it i}) sampling time interval $[t_i,t_f]$, {\it ii}) number of time steps $n_{\textrm{tstep}}$, and {\it iii}) the number of Trotter steps $n_{\textrm{trotter}}$.  
Then the size of matrices $\mathbf{S}=\{S_{ij}\}$ and $\mathbf{F}=\{F_{ij}\}$ is $n_{\textrm{tstep}}\times n_{\textrm{tstep}}$.
In practical simulations $\mathbf{S}$ is usually ill-conditioned. 
For numerical stability we applied the canonical orthogonalization scheme as a postprocessing step before the diagonalization.
The matrix size is then reduced to $n_{\textrm{cut}}\times n_{\textrm{cut}}$ where $n_{\textrm{cut}}$ is some integer satisfying $0<n_{\textrm{cut}} < n_{\textrm{tstep}}$. 
\subsection{Measurement of the matrix elements}
\label{subsec:measurement}
In this section we introduce a way to estimate the complex phase factor of the matrix element.
Mapping fermions to bosons changes the fundamental structure of the Hamiltonian.
The expectation value of boson number operator $\hat{n}=\sum_{\bm{Q}}\beta^\dagger_{\bm{Q}}\beta_{\bm{Q}}$ is no longer a conserved quantity.
It means the bosonic space may not possess a convenient subspace whose time evolution is trivial.
Cortes and Gray proposed the multi-fidelity estimation protocol which utilizes a zero particle state to make interference with the system in question \cite{Cortes_2022}.
We therefore need to adopt a different approach by adding auxiliary qubits where the system Hamiltonian now becomes
\begin{eqnarray}
    \hat{H}\rightarrow \hat{H}' = \hat{H} \bigotimes_i^{n_{\rm{aux}}}\frac{1}{2}(I_i-\sigma^z_i)
    \label{eq:def_Haux}
\end{eqnarray}
Here $I_i-\sigma^z_i$ acts on the {\it i} th auxiliary qubit.
Then we prepare the initial state $\ket{\phi_0}$ as
\begin{eqnarray}
    \ket{\phi_0}=\frac{1}{\sqrt{2}}(\ket{00\dots 0}_{\rm{aux}}+\ket{11\dots1}_{\rm{aux}}) \otimes\ket{0}^{\otimes N}.
\end{eqnarray}
Then it is clear that its time evolution by $t$ is
\begin{eqnarray}
    &&\ket{\phi(t)}=e^{-it\hat{H}'} \ket{\phi_0}\nonumber\\
    &&=\frac{1}{\sqrt{2}}(\ket{00\dots0}_{\rm{aux}}\otimes\ket{0}^{\otimes N}+\ket{11\dots1}_{\rm{aux}}\otimes e^{-it\hat{H}}\ket{0}^{\otimes N}). \nonumber \\
\end{eqnarray}
Taking the overlap with $\ket{\phi_0}$, we have
\begin{eqnarray}
    \braket{\phi_0|\phi(t)}=\frac{1}{2}\{1+\bra{0}^{\otimes N}e^{-it\hat{H}}\ket{0}^{\otimes N}\}.
\end{eqnarray}
We define its amplitude by 
\begin{eqnarray}
    R_1\equiv|\braket{\phi_0|\phi(\tau)}|
\end{eqnarray}
and
\begin{eqnarray}
    R_0 e^{i\theta_0}\equiv \bra{0}^{\otimes N}e^{-it\hat{H}}\ket{0}^{\otimes N}
\end{eqnarray}
where $R_0, R_1, \theta_0\in \mathbb{R}$. 
$\theta_0$ is what we desire to get.
It is then easily derived that
\begin{eqnarray}
    \cos\theta_0=\frac{4R_1^2-R_0^2-1}{2R_0}.
\end{eqnarray} 
$\sin\theta_0$ can be derived in the same manner by twisting the phase of an appropriate qubit by $\pi/2$.

\section{Numerical Experiment}
\label{sec:results}
We investigate the numerical performance of spinless boson representation for the ground state energy calculation of Hydrogen chains.
The crystal structure is given in Fig.~(\ref{fig:H_dimer}) (a).
The unit cell includes two hydrogen atoms and the interatomic distance $R$ is defined so that $R \le L/2$ where $L$ is the lattice constant.
We show the results for two cases, namely a dimerized ($R < L/2$) and an equidistant ($R = L/2$) hydrogen chain.
For a sufficiently large $n_k$, the number of $k$ point sampling, the former case is expected to be a band insulator and the latter is in a metallic or Mott insulator phase in its ground state.

\subsection{Dimerized Hydrogen Chain}
\label{subsec:H_dimer}
\begin{figure}[t]
  \raggedright
  \begin{minipage}[t]{0.45\linewidth}
    \vspace{3mm}
    \centering
    \begin{overpic}[width=\linewidth]{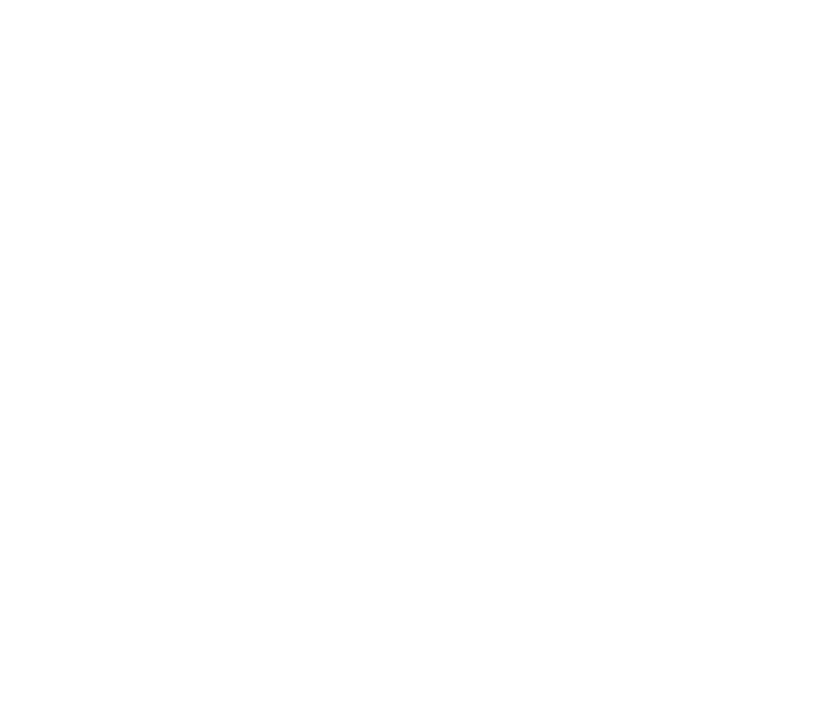}
    \put(0,15){
    \resizebox{\linewidth}{!}{\begin{tikzpicture}
    \draw[line width=0.3mm,-{Stealth[length=1mm,width=1mm]}] (0, 0) -- (3,0);
    \draw[line width=0.3mm, dotted] (0.3, -1) -- (0.3,0.6);
    \draw[line width=0.3mm, dotted] (2.4, -1) -- (2.4,0.6);
    \filldraw[fill=gray, draw=black, thick] (0.3,0) circle (0.2);
    \filldraw[fill=gray, draw=black, thick] (1.2,0) circle (0.2);
    \filldraw[fill=gray, draw=black, thick] (2.4,0) circle (0.2);
    \draw[line width=0.2mm,{Stealth[length=1mm,width=1mm]}-{Stealth[length=1mm,width=1mm]}] (0.3, -0.84) -- (2.4,-0.84);
    \draw[line width=0.2mm,{Stealth[length=1mm,width=1mm]}-{Stealth[length=1mm,width=1mm]}] (0.3, -0.45) -- (1.2,-0.45);

    \node[font=\fontsize{11}{24}\selectfont] at (3,-0.16) {$z$};
    \node[font=\fontsize{9}{24}\selectfont, text=white] at (0.3,0.0) {\ce{H}};
    \node[font=\fontsize{9}{24}\selectfont] at (0.75,-0.6) {$R$};
    \node[font=\fontsize{9}{24}\selectfont] at (1.35,-1.1) {$L$};

\end{tikzpicture}}
    }
    \put(0,85){\panelLabel{(a)}}
    \end{overpic}
  \end{minipage}
  \begin{minipage}[t]{0.45\linewidth}
    \vspace{3mm}
    \centering
    \begin{overpic}[width=\linewidth]{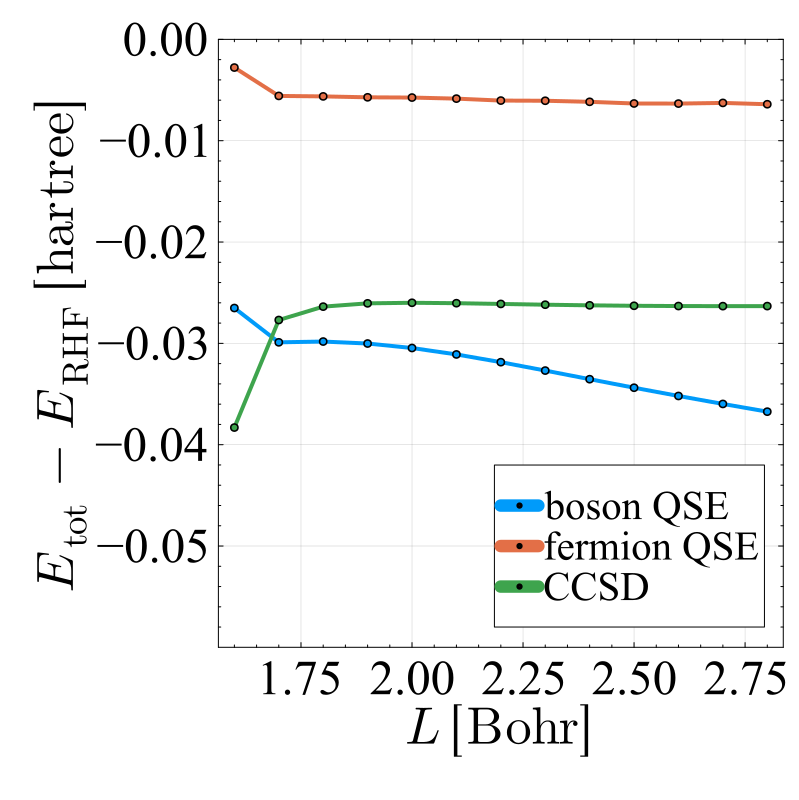}
      \put(0,100){\panelLabel{(b)}}
    \end{overpic}
  \end{minipage}

  \begin{minipage}[t]{0.45\linewidth}
    \vspace{3mm}
    \centering
    \begin{overpic}[width=\linewidth]{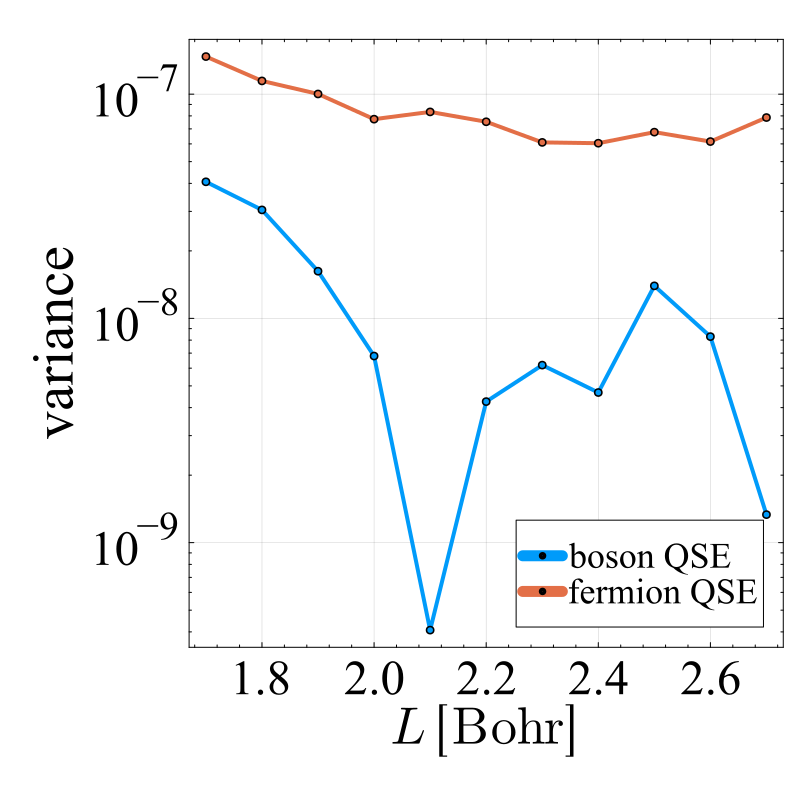}
      \put(0,100){\bfseries (c)}
    \end{overpic}
  \end{minipage}
  \begin{minipage}[t]{0.45\linewidth}
    \vspace{3mm}
    \centering
    \begin{overpic}[width=\linewidth]{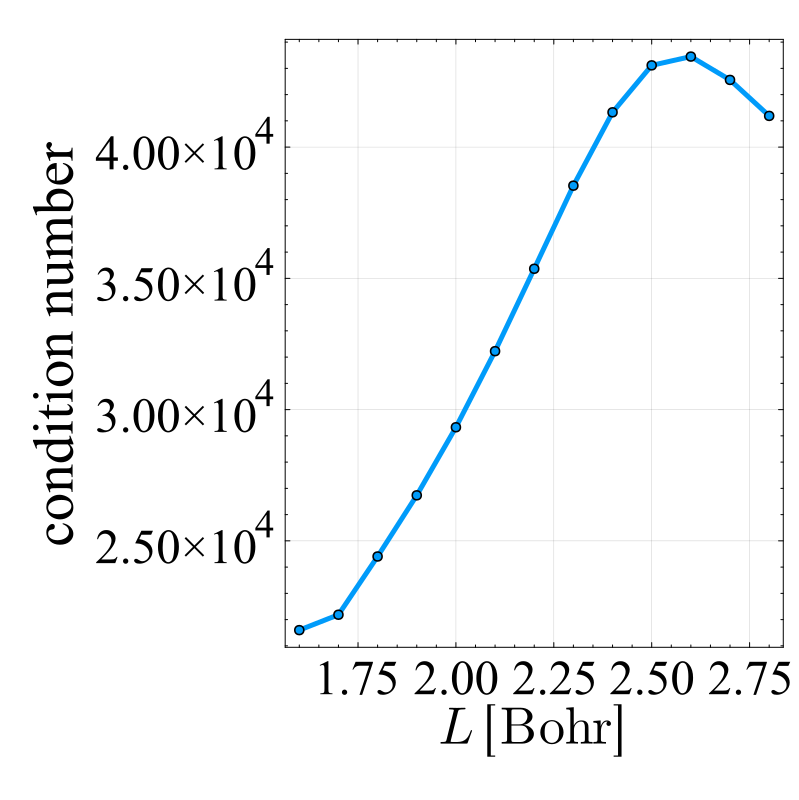}
      \put(0,100){\bfseries (d)}
    \end{overpic}
  \end{minipage}
  \begin{minipage}[t]{0.45\linewidth}
  \end{minipage}

  \caption{The crystal structure of a dimerized hydrogen chain is given in (a). $R$ is the interatomic distance and $L$ is the lattice constant.
  (b) Comparison of the spinless boson QSE and its fermionic counterpart for the ground state energy calculation of dimerized hydrogen chain measured from the energy of RHF.
  The results of CCSD calculations are given for reference as well. 
  The variance of operator Eq.~(\ref{eq:def_f}) is given for spinless boson QSE and spinless fermion QSE in (c).
  (d)The condition number of Gram matrix $\mathbf{S}$ in Eq.~(\ref{eq:def_Sij}) , whose dimension is reduced to $n_{\textrm{cut}}=2$ by canonical orthogonalization.} 
  \label{fig:H_dimer}
\end{figure}
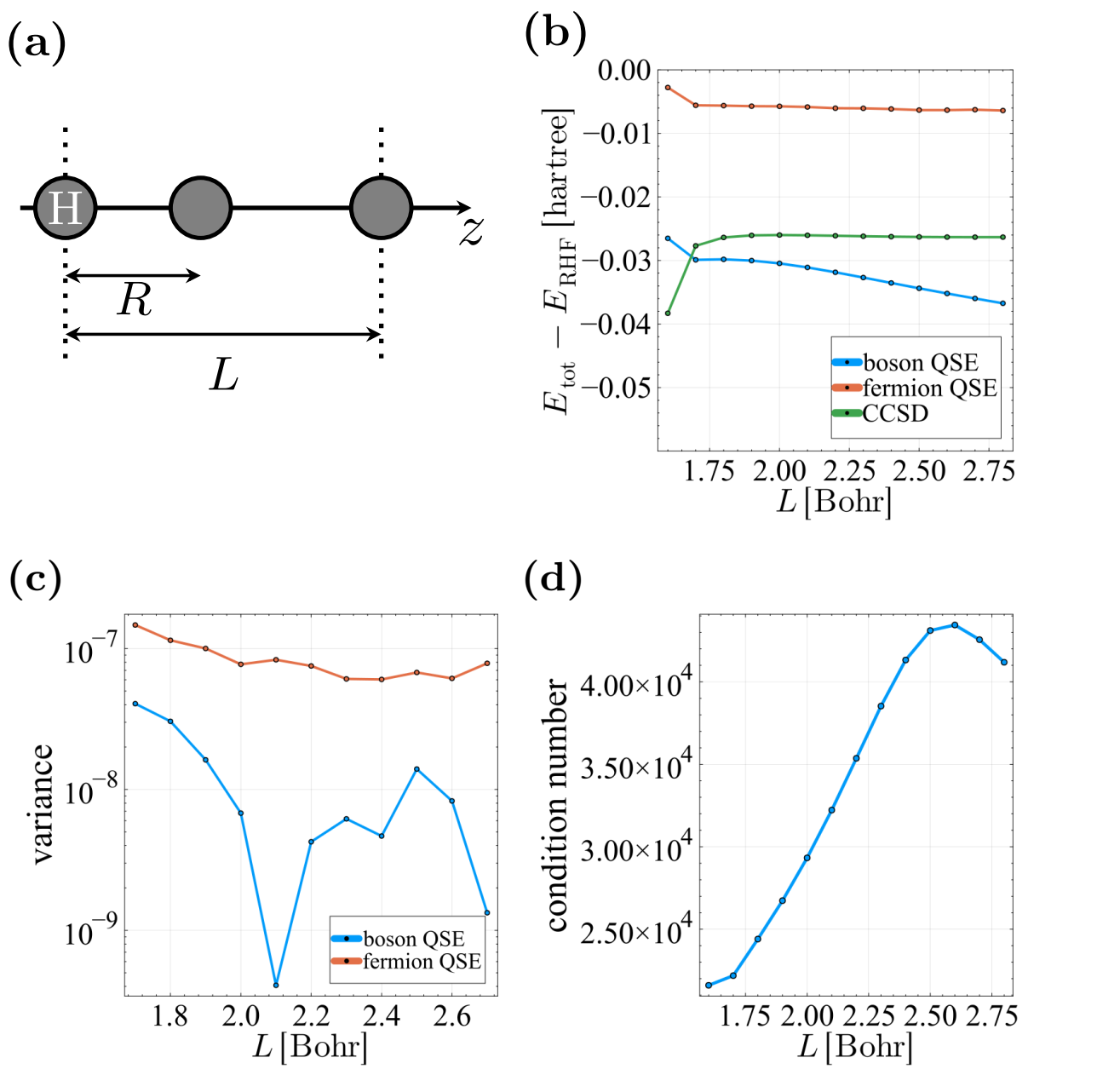

In Fig.~\ref{fig:H_dimer} (b) we show the ground state energy curve of the dimerized hydrogen chain.
In this calculation we employed the number of $k$ points $n_k=2$.
The number of Trotter steps $n_{\textrm{trotter}}$ is $n_{\textrm{trotter}}=20$.
We sampled six equally spaced points over the time interval $[0,0.6]$ in atomic units.  
Then the problem size is reduced to $2\times2$ matrix after canonical orthogonalization. 
We tried $n_{\textrm{cut}} > 2$ though such a setup provided unphysical potential energy curves whose variance is orders of magnitude larger than that obtained with $n_{\textrm{cut}}=2$. 
The reference state is the restricted Hartree-Fock (RHF) ground state.
For a fixed value $R=0.8$ Bohr, $L$ is varied from $0.8$ to $2.8$ Bohr.
A result by CCSD calculation is given for comparison. 
According to the work by Manrique {\it et al.}~the deviation of CCSD from FCI is expected to be sufficiently small \cite{Manrique_2020}.   
The CCSD energy can thus be a scale to measure the error induced by approximations.

In Fig.~\ref{fig:H_dimer} (b) the spinless boson QSE deviates from CCSD by $2.2\times10^{-3}$ hartree at $L=1.7$ Bohr, which is the minimum error through this parameter range. 
Large $L$ scales up the error which amounts to $1.0\times 10^{-2}$ hartree at $L=2.8$ Bohr.
The spinless boson representation shows fairly good performance in spite of the two approximations: discarding spin degrees of freedom and bosonizing electron-hole pairs.
We disentangle the effect of two approximations.
We plot the results of QSE based on ordinary fermionic representation while the spin is discarded (spinless fermion QSE).
Except at the equidistant configuration of atoms at $L=1.6$, the spinless fermion QSE overestimates the energy by 0.02 hartree.
This is the error caused by discarding the spin. 
We, in turn, compare the two QSE methods.
The spinless boson QSE provides a deeper energy by $2.4\times 10^{-2}$ hartree at $L=1.6$ and $3.0\times 10^{-2}$ hartree at $L=2.8$.
This is the error caused by bosonization.
We can conclude that the small error of spinless boson QSE is the consequence of error cancellation between the two appriximations.

In Fig.~\ref{fig:H_dimer} (c) we show the variance of operator Eq.~(\ref{eq:def_f}) corresponding to Fig.~\ref{fig:H_dimer} (b).
This is a criterion to judge the quality of solutions.
If the solution is exact its variance should always be zero.
The spinless boson QSE shows smaller variance through all values of $L$.
Figure~\ref{fig:H_dimer} (c) indicates the variance of QSE is sensitive to $L$, hence corresponding electronic structure, as well as the particle's statistics.
To our knowledge the relevance of variance to these factors is not well investigated and we have no clear explanation at this moment.

In Fig.~\ref{fig:H_dimer} (d) we show the condition number of Gram matrix with $n_{\textrm{cut}}=2$.
The condition number is defined by the ratio of the maximum to the minimum eigenvalue of the Gram matrix $\mathbf{S}$.
A smaller value is preferred for the numerical stability of diagonalization.  
The condition number is $L$ dependent as well.
We note that the value is orders of magnitude smaller than the values reported in ref.~\cite{Cortes_2022}.
That is mostly because $n_{\textrm{cut}}$ is much smaller than the corresponding value in ref.~\cite{Cortes_2022}, which is 60 at most.
We found numerical instability of the energy curve plot for $n_{\textrm{cut}} \ge 3$ as we stated above.
Probably the reason is our way of the partition of Hamiltonian in Eq.~(\ref{eq:def_Hamiltonian}).
We expect $\Delta \hat{H}$ is perturbative, which indicates the mixture of higher energy states is suppressed and the Gram matrix tends to be more ill-conditioned. 

\begin{figure}[t]
  \begin{minipage}[t]{1\linewidth}
    \vspace{0pt}
    \centering
    
    \begin{minipage}{0.45\linewidth}
      \centering
      \begin{overpic}[width=\linewidth]{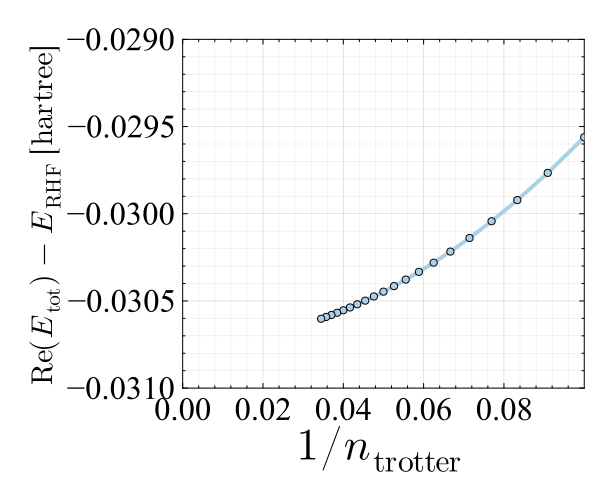}
        \put(0,85){\bfseries (a)}
      \end{overpic}
    \end{minipage}
    \begin{minipage}{0.45\linewidth}
      \centering
      \begin{overpic}[width=\linewidth]{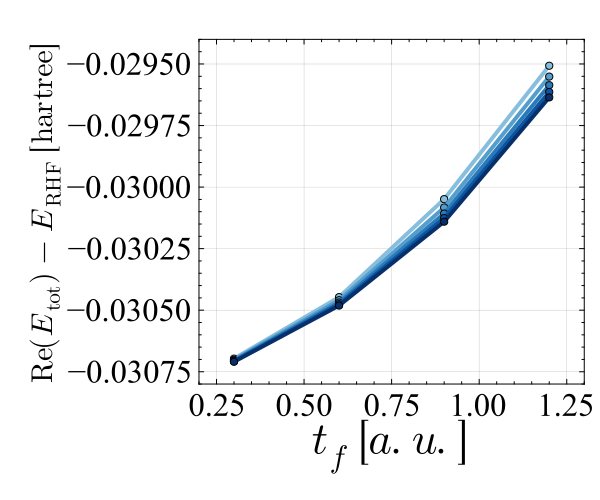}
        \put(0,85){\bfseries (d)}
      \end{overpic}
    \end{minipage}

    \begin{minipage}{0.45\linewidth}
      \centering
      \begin{overpic}[width=\linewidth]{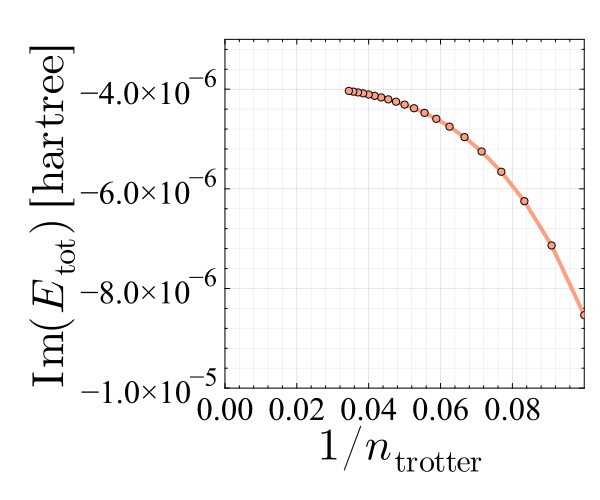}
        \put(0,85){\bfseries (b)}
      \end{overpic}
    \end{minipage}
    \begin{minipage}{0.45\linewidth}
      \centering
      \begin{overpic}[width=\linewidth]{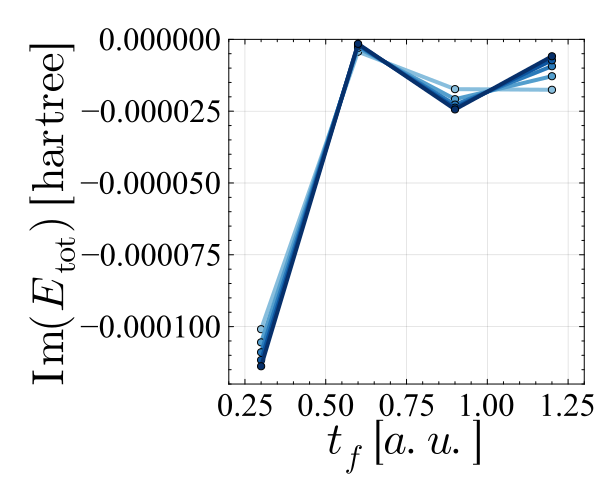}
        \put(0,85){\bfseries (e)}
      \end{overpic}
    \end{minipage}

    \begin{minipage}{0.45\linewidth}
      \centering
      \begin{overpic}[width=\linewidth]{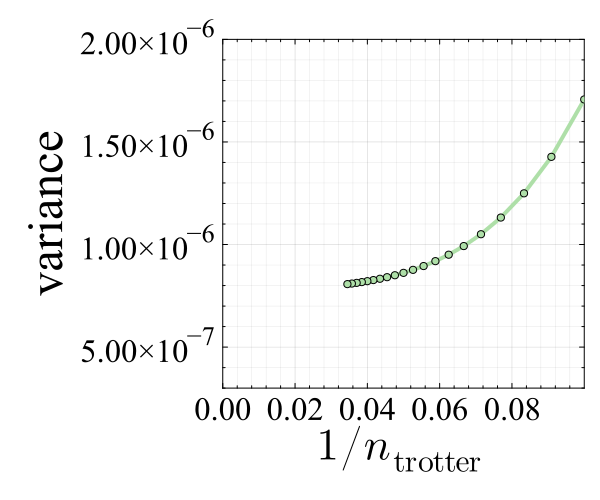}
        \put(0,85){\bfseries (c)}
      \end{overpic}
    \end{minipage}
    \begin{minipage}{0.45\linewidth}
      \centering
      \begin{overpic}[width=\linewidth]{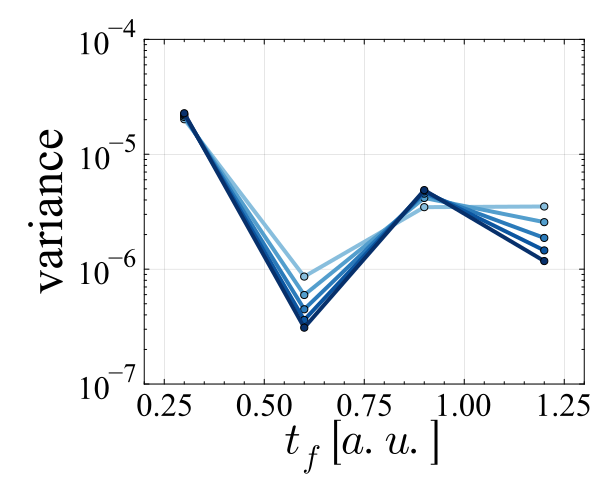}
        \put(0,85){\bfseries (f)}
      \end{overpic}
    \end{minipage}
  \end{minipage}
  \caption{Numerical behavior of spinless boson QSE is investigated for three parameters: {\it i})~the number of Trotter steps $n_{\textrm{trotter}}$, {\it ii})~time interval $[0,t_f]$, and {\it iii})~the number of sampling points $n_{\textrm{tstep}}$.
  The real and imaginary parts of the eigenvalues are shown for varying $n_{\textrm{trotter}}$ in (a)-(b) and $t_f$ in (d)-(e).
  Corresponding variance is given in (c) and (f). 
  $n_{\textrm{cut}}$ is the matrix dimension of Gram matrix $\mathbf{S}$ reduced by the canonical orthogonalization.
  In (d)-(f) $n_{\textrm{tstep}}=5$ for the lightest blue line and $n_{\textrm{tstep}}=10$ for the darkest.
  A fixed value $n_{\textrm{cut}}=3$ is used.}
  \label{fig:dimer_convergence}
\end{figure}

We examine the numerical convergence of three parameters introduced in the last section. 
In Fig.~\ref{fig:dimer_convergence} (a)-(c) we show the real and imaginary parts of total energy $E_{\textrm{tot}}$ and the variance of operator Eq.~(\ref{eq:def_f}).
We compare the convergence of three quantities with respect to the number of Trotter steps $n_{\textrm{trotter}}$ from $10$ to $30$.
$n_{\textrm{cut}}=2$ through all the calculations in Fig.~\ref{fig:dimer_convergence}.
Here $R=0.8$ Bohr and $L=2.0$ Bohr.
In Fig.~\ref{fig:dimer_convergence} (a) we see the monotonic improvement of energy with the increase of $n_{\textrm{trotter}}$.
As shown in Fig.~\ref{fig:dimer_convergence} (b) the QSE eigenvalues have non-negligible imaginary parts as a numerical error.
It is probably because the operator Eq.~(\ref{eq:def_f}) is not Hermitian and finite $n_{\textrm{trotter}}$ causes numerical errors that lead to a nonzero imaginary part.
Although the magnitude is sufficiently small, Fig.~\ref{fig:dimer_convergence} (b) shows a slow convergence of the imaginary part.
The variance of operator Eq.~(\ref{eq:def_f}) is given in Fig.~\ref{fig:dimer_convergence} (c).
We expect the variance vanishes in the $1/n_{\textrm{tstep}} \to 0$ limit but the convergence is slow.

In Fig.~\ref{fig:dimer_convergence} (d)-(f) we vary the range of sampling time for $t\in [0,t_f]$ and the number of sampling points $n_{\textrm{tstep}}$.
Here $n_{\textrm{trotter}}=20$ and $n_{\textrm{cut}}=2$.
As expected from the variational principle, at a given $t_f$, larger $n_{\textrm{tstep}}$ always leads to energy improvement. 
See Fig.~\ref{fig:dimer_convergence} (d) for this point.
In contrast to the real part of energy, larger $t_f$ and $n_{\textrm{tstep}}$ do not necessarily reduce the imaginary part of energy and the variance, as given in Fig.~\ref{fig:dimer_convergence} (e) and (f).

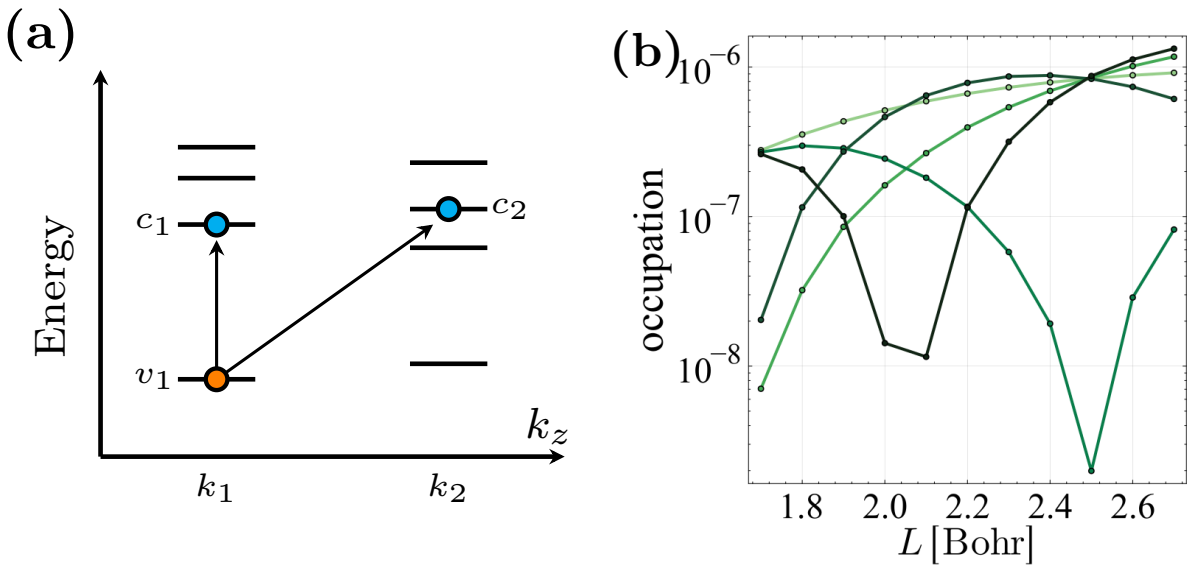
\begin{figure}[t]

\begin{minipage}[t]{0.45\linewidth}
\vspace{3mm}
\centering
\begin{overpic}[width=\linewidth]{./Figures/transparent.png}
\put(0,0){
\resizebox{\linewidth}{!}{\begin{tikzpicture}
    \draw[line width=0.3mm,-{Stealth[length=1mm,width=1mm]}] (0, 0) -- (0,2.5);
    \draw[line width=0.3mm,-{Stealth[length=1mm,width=1mm]}] (0, 0) -- (3,0);
    \node[font=\fontsize{8}{24}\selectfont] at (2.9,0.2) {$k_z$};
    \node[font=\fontsize{8}{24}\selectfont, rotate=90] at (-0.3,1.1) {Energy};

    \draw[line width=0.3mm] (0.5, 0.5) -- (1.0,0.5);
    \draw[line width=0.3mm] (0.5, 1.5) -- (1.0,1.5);
    \draw[line width=0.3mm] (0.5, 1.8) -- (1.0,1.8);
    \draw[line width=0.3mm] (0.5, 2.0) -- (1.0,2.0);

    \draw[line width=0.3mm] (2.0, 0.6) -- (2.5,0.6);
    \draw[line width=0.3mm] (2.0, 1.35) -- (2.5,1.35);
    \draw[line width=0.3mm] (2.0, 1.6) -- (2.5,1.6);
    \draw[line width=0.3mm] (2.0, 1.9) -- (2.5,1.9);

    \draw[line width=0.2mm,-{Stealth[length=1mm,width=1mm]}] (0.75,0.5) -- (0.75,1.4);
    \draw[line width=0.2mm,-{Stealth[length=1mm,width=1mm]}] (0.75,0.5) -- (2.15,1.5);
    \filldraw[fill=orange, draw=black, thick] (0.75,0.5) circle (0.07);    
    \filldraw[fill=cyan, draw=black, thick] (0.75,1.5) circle (0.07);
    \filldraw[fill=cyan, draw=black, thick] (2.25,1.6) circle (0.07);
    \node[font=\fontsize{6}{24}\selectfont] at (0.35,0.5) {$v_1$};
    \node[font=\fontsize{6}{24}\selectfont] at (0.35,1.5) {$c_1$};
    \node[font=\fontsize{6}{24}\selectfont] at (2.65,1.6) {$c_2$};
    \node[font=\fontsize{6}{24}\selectfont] at (0.75,-0.2) {$k_1$};
    \node[font=\fontsize{6}{24}\selectfont] at (2.25,-0.2) {$k_2$};
\end{tikzpicture}}
}
\put(0,80){\panelLabel{(a)}}
\end{overpic}
  \end{minipage}
  \begin{minipage}[t]{0.45\linewidth}
    \vspace{3mm}
    \centering
    \begin{overpic}[width=\linewidth]{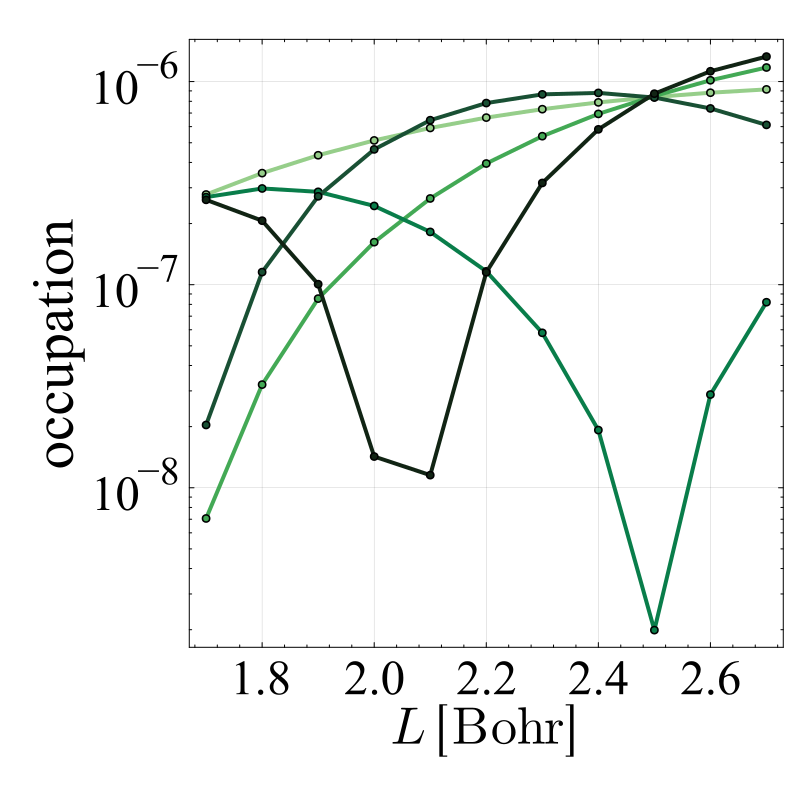}
      \put(0,90){\bfseries (b)}
    \end{overpic}
  \end{minipage}

  \caption{(a) A schematic picture of a doubly occupied state Eq.~(\ref{eq:def_double_occupation}). (b) The time evolution of the amplitude of doubly occupied state is plotted in. 
  The amplitude is plotted for five time steps $t=0.12, 0.24 ,0.36 ,0.48 ,0.6$ a.u..
  The earlier the time step is, the brighter the line color is. The lightest green curve is at $t=0.12$. }
  \label{fig:dimer_occupation}
\end{figure}

Now we examine an issue inherent to the bosonization.
Although we truncate the highest occupation number of each bosonic mode in boson-qubit mapping, this does not prohibit the following type of unphysical occupied state:
\begin{eqnarray}
\beta_{c_1v_1}^\dagger\beta_{c_2v_1}^\dagger\ket{0},
\label{eq:def_double_occupation}
\end{eqnarray}
where the index $v_1$ denotes  a valence band and $c_1$ and $c_2$ ($c_1\neq c_2$) denote conduction bands.
A schematic picture of this process is given in Fig.~\ref{fig:dimer_occupation} (a).
It should be prohibited in the fermionic system.   
The $L$ dependency and time evolution of the amplitude of Eq.~(\ref{eq:def_double_occupation}) are plotted in Fig.~\ref{fig:dimer_occupation} (b).
It shows non-monotonic and complex behavior but it is bounded by less than $10^{-6}$ in this parameter range, whose energetic contribution is expected to be negligible. 

\subsection{Equidistant Hydrogen Chain}
\label{subsec:H_equidistant}

\begin{figure}[t]
  \raggedright

  \begin{minipage}{0.47\linewidth}
    \begin{overpic}[width=\linewidth]{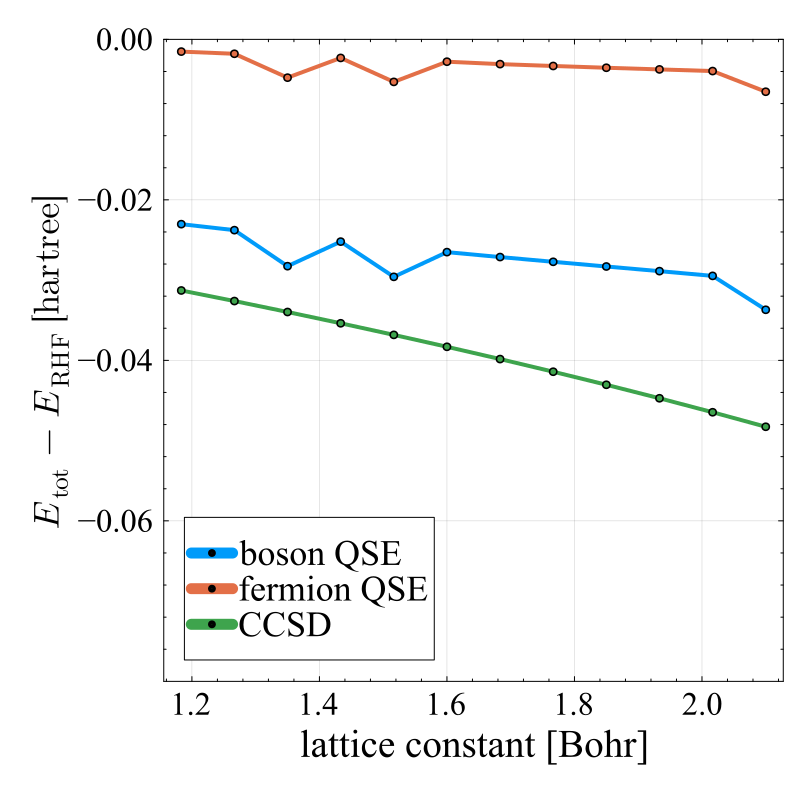}
      \put(0,100){\bfseries (a)}
    \end{overpic}
  \end{minipage}
  \begin{minipage}{0.47\linewidth}
    \begin{overpic}[width=\linewidth]{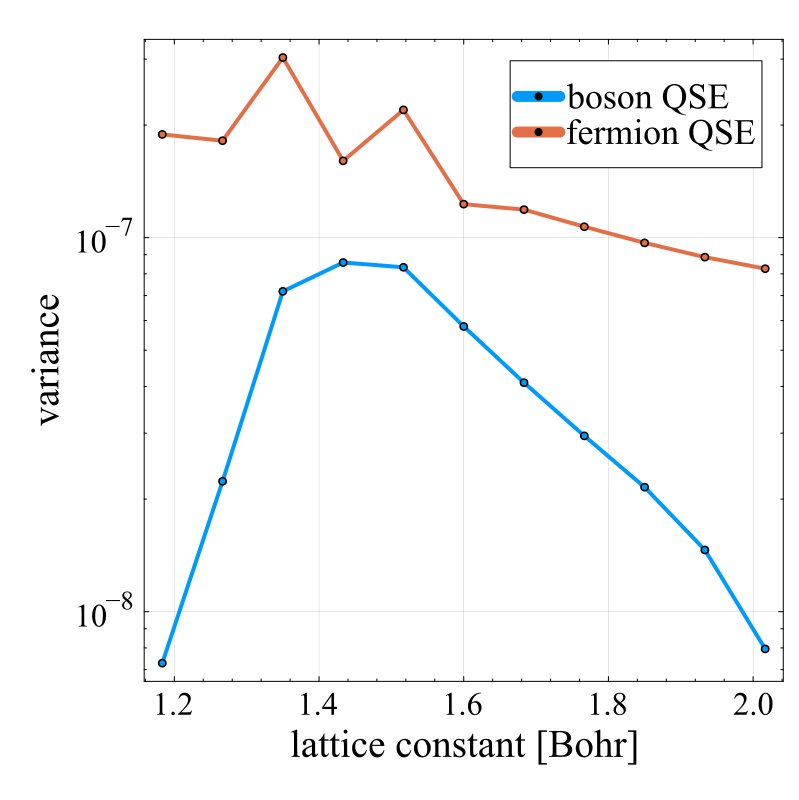}
      \put(0,100){\bfseries (b)}
    \end{overpic}
  \end{minipage}

  \begin{minipage}{0.47\linewidth}
    \begin{overpic}[width=\linewidth]{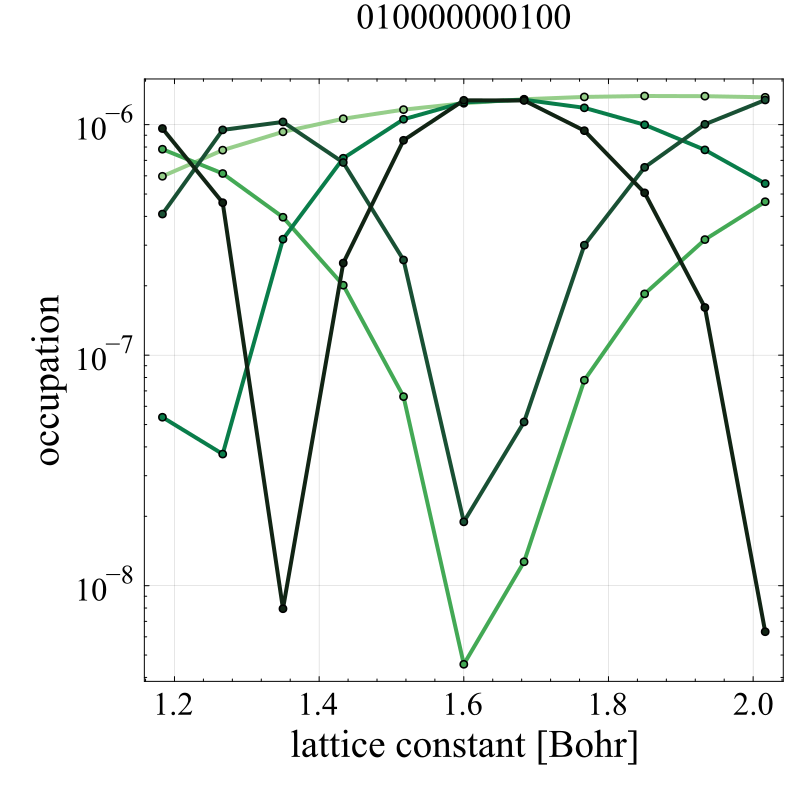}
      \put(0,90){\bfseries (c)}
    \end{overpic}
  \end{minipage}
  \begin{minipage}{0.47\linewidth}
    \begin{overpic}[width=\linewidth]{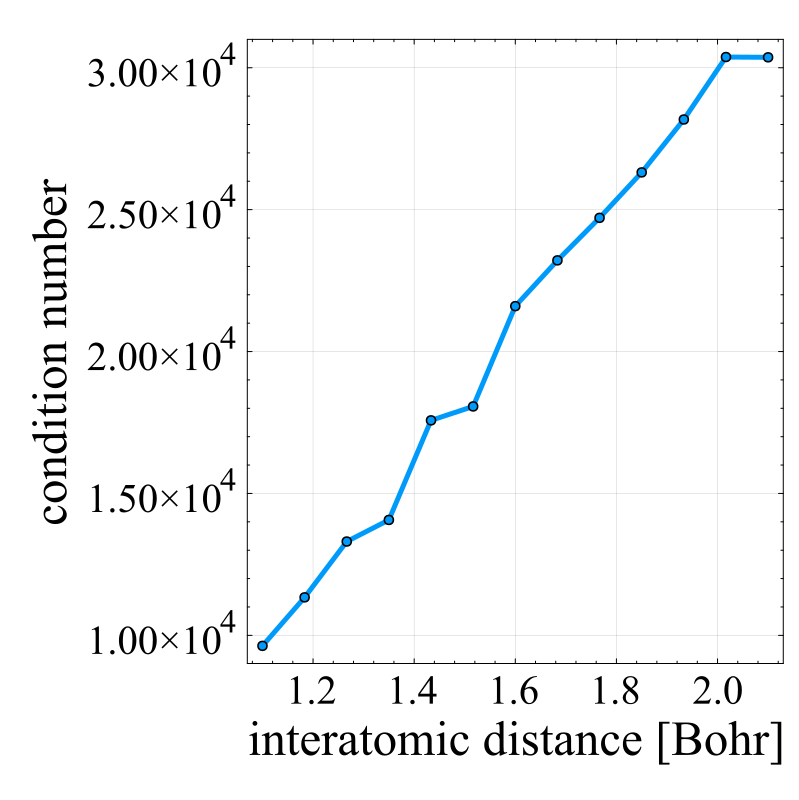}
      \put(0,90){\bfseries (d)}
    \end{overpic}
  \end{minipage}

  \caption{The ground state energy of equidistant hydrogen chain measured from the ground state energy of restricted Hartree-Fock (RHF) calculation. We show calculations by spinless boson QSE (blue line) and spinless fermion QSE (orange line) (a). 
  The results of CCSD calculation are also plotted for comparison (green line). The variance of operator Eq.~(\ref{eq:def_f}) is in (b) for spinless boson QSE (blue line) and spinless fermion QSE (orange line) on a linear scale.
  (c) Time evolution of the amplitude of doubly occupied state for five time steps $t=0.12, 0.24 ,0.36 ,0.48 ,0.6$ a.u..
  The earlier the time step is, the brighter the line color is. The lightest green curve is at $t=0.12$.
  The condition number of Gram matrix $\mathbf{S}$ is plotted in (d).}
  \label{fig:H_equidistant}
\end{figure}

In Fig.~\ref{fig:H_equidistant} (a) we show the ground state energy of an equidistant hydrogen chain ($R=L/2$) simulated by spinless boson QSE varying the value of $L$.
The energy is measured from the ground state energy of RHF.
The number of $k$ point $n_k$ is 2. 
QSE parameters are $n_{\textrm{tstep}}=6$, $n_{\textrm{cut}}=2$, and $n_{\textrm{trotter}}=20$.
In the thermodynamic limit $n_k\to \infty$ the ground state of equidistant hydrogen chain is predicted to be a metal for $L < 1.6$ and a Mott insulator for $L > 1.6$ \cite{Motta_2020}.
While running the QSE on circuit simulators with large $n_k$ is not computationally affordable, it is tempting to test the numerical behaviour in this parameter region.
We again compare the result with spinless fermion QSE and CCSD methods.
We see in Fig.~\ref{fig:H_equidistant} (a) that the energetical hierarchy of three methods is similar to the case of dimerized Hydrogen chain in Fig.~\ref{fig:H_dimer}.
The energy gap between spinless fermion QSE and CCSD mostly owes to the approximation of discarding spin degrees of freedom.
It is plausible that restricting the variational space leads to overestimation of the ground state total energy. 
The spinless boson QSE again predicts a lower energy than the spinless fermion QSE by 0.03 hartree at maximum in this parameter range. 
The error cancellation between the two approximations results in a small overestimation of the energy of spinless boson QSE with respect to CCSD.
The overestimation is $8.3\times 10^{-3}$ hartree at $L=1.18$ and $1.45\times 10^{-2}$ hartree at $L=2.1$.

In Fig.~\ref{fig:H_equidistant} (b) we plot the variance of operator Eq.~(\ref{eq:def_f}).
Now the spinless boson QSE shows a smaller variance than spinless fermion QSE like we found in the dimerized case.
Figure \ref{fig:H_equidistant} (c) shows the amplitude of doubly occupied state Eq.~(\ref{eq:def_double_occupation}).
While the dynamics is complex and strongly $L$ dependent, the amplitude is bounded by around $10^{-6}$ and is negligible.
We lastly plot the condition number in Fig.~\ref{fig:H_equidistant} (d).
Despite the difference of expected thermodynamic limits of ground state, 
the numerical behaviour of the equidistant hydrogen chain resembles that of the dimerized system.
For a more concrete diagnosis of the many-body phases the implementation of polarization operator is necessary as discussed in ref.~\cite{Motta_2020}. 
This will be addressed in future work.

\section{Discussion}
\label{sec:discussion}

In this work we introduced a boson representation of electron-hole pairs for its potential to compress the quantum circuit depth.
We observed that the spinless boson QSE shows practical accuracy in the ground state energy calculations of periodic Hydrogen chains.
Here we discuss further details to explain the numerical aspects of approximations we introduced.
We found the instability of QSE for $n_{\textrm{cut}} > 3$, which means the overcompleteness of the Gram matrix arises at a much smaller value of $n_{\textrm{cut}}$ than previously reported \cite{Cortes_2022}.  
Considering our observation that the same problem appears in spinless fermion QSE, it may originate from the splitting of Hamiltonian Eq.~(\ref{eq:def_Hamiltonian}).
Contrary to the bare Hamiltonian in \cite{Cortes_2022} the large matrix elements are renormalized into Hartree-Fock orbital energy $\epsilon_{km}$ or largely cancelled out in Eq.~(\ref{eq:def_dH}).
Consequently the Krylov subspace spanned by the exponential of the Hamiltonian in Eq.~(\ref{eq:def_Hamiltonian}) can be smaller than that spanned by using the bare Hamiltonian.
In such a case the overcompleteness may arise at a smaller value of $n_{\textrm{cut}}$.

An important question is the applicability of boson representation in higher dimensional systems.
It is computationally challenging to examine two- and three-dimensional systems on circuit simulators,
yet we expect the precision will be improved.  
It is because the Pauli exclusion principle is pronounced when the particles are confined over short distances.
In such a case the energy level spacing is large and bosonic particles tend to occupy identical orbitals
which may cause serious unphysical double occupations.
In that sence the confinement is weaker than that in a hydrogen chain in more-than-two dimensional systems and will provide better precision.

Reduction of the total number of qubits is an important issue. 
We introduced the boson representation motivated by a possible circuit depth suppression.
One issue is that it requires a much larger number of qubits than the original fermionic description does,
because we have to prepare all possible combinations of electron-hole pairs.
We think the reduction is possible by focusing on a redundancy of the present formalism.
As an empirical fact a major contribution of the self energy correction comes from the screening of long range part of Coulomb potential \cite{Rasmussen_2016, da_Jornada_2017,R_dl_2017}.
It indicates we can discard unimportant electron-hole pairs with large momentum transfer.

It is worth considering ways to refine the ansatz of boson representation.
Our bosonization extends the Hilbert space so that it involves unphysical states like doubly occupied states.
This artifact may not be negligible in molecular systems.
The unphysical states can be excluded by projecting the system onto an antisymmetrized subspace by applying the Marumori operator.
See Appendix \ref{sec:antisymmetrization}.
The antisymmetrization, however, requires a superposition of all bosonic states.
It spoils our original intention to compress circuit depth.
Another possible way would be introducing an artificial repulsive potential to avoid multiple occupation.
It means an electron-hole pair boson is treated as a hard-core boson \cite{Matsubara_1956}.
In that case the potential depth is a tunable parameter and the framework is not {\it ab initio} anymore.
We lastly comment on its applicability to the excited states.
We need to explicitly involve the spin degrees of freedom for those cases.  
If one focuses on the energetically low lying states of band insulators, an electron-hole composite boson formulation can be a more natural choice since such states are dominated by excitonic many-body states.
Such an approach has a decades-long history \cite{Combescot_2007,Noordman_2026} and may provide insight into the design of more efficient formulations.

\section{Concluding Remarks}
\label{sec:concluding_remarks}

Motivated by the development of quantum algorithms for first-principles simulations of solid-state electronic systems, 
we proposed a bosonic reperesentation in which electron-hole pair excitations are treated as bosonic degrees of freedom.
This reperesentation is expected to be justified when the true many-electron state can be described by a perturbative correction 
to the electronic state obtained within the mean-field approximation.
By adopting the bosonic reperesentation, the nonlocal Pauli-string operators arising from the Jordan-Wigner transformation of the original fermionic system can be eliminated,
 thereby reducing the depth of quantum circuits.
In this work, we investigated the quantitative predictive power of such a bosonic representation and examined its numerical behavior.
To compute the ground-state energies of electronic systems, we employed the Quantum Subspace Expansion (QSE) method.
Since QSE does not require gradient evaluations and is generally considered more robust against noise than the Variational Quantum Eigensolver (VQE), it is regarded as a promising approach for applications on noisy intermediate-scale quantum (NISQ) devices.
The QSE solver was implemented using Qulacs\cite{Suzuki_2021}, OpenFermion\cite{McClean_2020}, and Qiskit\cite{Javadi-Abhari_2024}.
The restricted Hartree-Fock (RHF) reference states were obtained using PySCF\cite{Sun_2017}.

Numerical calculations were performed for the ground-state energies of the dimerized and equidistant hydrogen chains under periodic boundary conditions.
These two cases show different many-body phases in the thermodynamic limit and hence serve as an ideal test bed for our algorithm.
We confirmed that the boson reperesentation realizes practical accuracy for ground-state energy calculations of both cases. 
We also systematically investigated the numerical aspects of the convergence with respect to parameters of QSE. 
In the canonical orthogonalization procedure, no more than three basis states could be retained reliably, 
whereas enlarging the subspace often resulted in numerical instabilities.
It might be explained by the representation of the Hamiltonian as the sum of a mean-field term and a perturbation. 
We believe that these findings provide strong evidence for the usefulness of the bosonic reperesentation in quasi-two-dimensional and three-dimensional systems, including realistic condensed-matter materials.
Nevertheless, reducing the number of bosonic modes to improve computational efficiency and achieving a deeper understanding of the connection between the bosonic representation and the numerical properties of QSE remain important subjects for future investigation.

\appendix

\section{Quantum circuits of the boson representation}
\label{sec:hamiltonian}
In this section we provide a complete list of the boson Hamiltonian for $\Delta\hat{H}$ in Eq.~(\ref{eq:def_dH}).
We use shorthand notation for the indices $\mathbf{v} = (\bm{k},v)$ and $\mathbf{c} = (\bm{p},c)$.
The electron repulsion integral (ERI) or matrix element of Coulomb interaction in physicists’ notation is given by
\begin{eqnarray}
    &&w_{\bm{k}_1m_1\bm{k}_2m_2\bm{k}_3m_3\bm{k}_4m_4}\nonumber \\
    &&=\frac{1}{2}\int d^3\bm{r}\phi_{\bm{k}_1m_1}(\bm{r}_1)^*\phi_{\bm{k}_2m_2}(\bm{r}_2)^*\frac{1}{r_{12}}\phi_{\bm{k}_3m_3}(\bm{r}_2)\phi_{\bm{k}_4m_4}(\bm{r}_1).\nonumber \\
    \label{eq:def_ERI}
\end{eqnarray}
$\phi_{\bm{k}m}(\bm{r})$ is the one-particle orbital of crystal momentum $\bm{k}$ and band index $m$.
Equation (\ref{eq:def_ERI}) vanishes when the crystal momentum conservation is not satisfied.
$w_{\rm{mf}}^v$ and $w_{\rm{mf}}^c$ are mean-field contributions given by
\begin{eqnarray}
    &&w_{\rm{mf};\bm{k}_1m_1\bm{k}_2m_2}^v\nonumber\\
    &&=\sum_{\bm{k},v}(n_s^2w_{\bm{k}_1m_1\bm{k}v\bm{k}v\bm{k}_2m_2}-n_sw_{\bm{k}_1m_1\bm{k}v\bm{k}_2m_2\bm{k}v})\\
    &&w_{\rm{mf};m_1m_2}^c\nonumber\\
    &&=\sum_{\bm{p},c}(n_s^2w_{\bm{k}_1m_1\bm{p}c\bm{p}c\bm{k}_2m_2}-n_sw_{\bm{k}_1m_1\bm{p}c\bm{k}_2m_2\bm{p}c}).
\end{eqnarray}
``$\rm{h. c.}$'' means taking Hermitian conjugate of terms prior to this symbol.
We classify every term of boson Hamiltonian into operators acting on $n$ qubits ($n=1,2,3,4$).
For each $n$ we show a matrix product form of Hermitian operators $\hat{h}_{n;i}$.
Corresponding boson operators are summarized in tables.
The  ``index'' column means the bands to which ERI's orbitals belong.
The index ``vvcv'' means that the corresponding term originates from $w_{\bm{v}_i\bm{v}_j\bm{c}_i\bm{v}_k}$. 
Now we express $\Delta\hat{H}$ by
\begin{eqnarray}
    \Delta H = \sum_{n,i}\hat{h}_{n;i} - \hat{V}_{\textrm{HF}},
\end{eqnarray}
where the first sum only includes the bare Coulomb interaction.
The index $n$ indicates the number of qubits it acts on, and $i$ indicates the type of operator.
$\hat{V}_{\textrm{HF}}$ is the sum of the Hartree term and Fock term.
\subsection{one-qubit operator}
\begin{itemize}

    \item $\hat{h}_{1;1}=h\begin{bmatrix} 0 & 0 \\ 0 & 1 \end{bmatrix}_{\bm{Q}}$
    \\
    \begin{center}
    \begin{tabular}{lc}
        \toprule
        index  & operator \\
        \midrule
        \parbox{0.075\textwidth}{vvvv} & \parbox{0.35\textwidth}{$\frac{1}{n_kn_c}w_{\rm{mf};\bold{v}_1\bold{v}_1}^v\beta_{\bold{c}_1\bold{v}_1}\beta_{\bold{c}_1\bold{v}_1}^\dagger$} \\
        \parbox{0.075\textwidth}{vcvc\\} & \parbox{0.35\textwidth}{$(-\frac{n_s}{n_kn_c}w_{\bold{v}_1\bold{c}_2\bold{v}_1\bold{c}_2}+n_sw_{\bold{v}_1\bold{c}_1\bold{v}_1\bold{c}_1})$\\$\times\beta_{\bold{c}_1\bold{v}_1}\beta_{\bold{c}_1\bold{v}_1}^\dagger$} \\
        \parbox{0.075\textwidth}{vccv\\} & \parbox{0.35\textwidth}{$(\frac{n_s^2}{n_kn_c}w_{\bold{v}_1\bold{c}_2\bold{c}_2\bold{v}_1} -n_sw_{\bold{v}_1\bold{c}_1\bold{c}_1\bold{v}_1})$\\$\times\beta_{\bold{c}_1\bold{v}_1}\beta_{\bold{c}_1\bold{v}_1}^\dagger$}\\
        \bottomrule
    \end{tabular}
    \end{center}
    
    \item $\hat{h}_{1;2}=h\begin{bmatrix} 1 & 0 \\ 0 & 0\end{bmatrix}_{\bm{Q}}$
    \\
    \begin{center}
    \begin{tabular}{lc}
        \toprule
        index & operator \\
        \midrule
        \parbox{0.075\textwidth}{cvcv\\} & \parbox{0.35\textwidth}{$(-\frac{n_s}{n_kn_v}w_{\bold{c}_1\bold{v}_2\bold{c}_1\bold{v}_2}+n_sw_{\bold{c}_1\bold{v}_1\bold{c}_1\bold{v}_1})$\\$\times\beta_{\bold{c}_1\bold{v}_1}^\dagger\beta_{\bold{c}_1\bold{v}_1}$}\\
        \parbox{0.075\textwidth}{cvvc} & \parbox{0.35\textwidth}{$(\frac{n_s^2}{n_kn_v}w_{\bold{c}_1\bold{v}_2\bold{v}_2\bold{c}_1}-n_sw_{\bold{c}_1\bold{v}_1\bold{v}_1\bold{c}_1})\beta_{\bold{c}_1\bold{v}_1}^\dagger\beta_{\bold{c}_1\bold{v}_1}$}\\
        \parbox{0.075\textwidth}{cccc} & \parbox{0.35\textwidth}{$\frac{1}{n_kn_v}w^c_{\rm{mf};\bold{c}_1\bold{c}_1}\beta^\dagger_{\bold{c}_1\bold{v}_1}\beta_{\bold{c}_1\bold{v}_1}$}\\
        \bottomrule
    \end{tabular}
    \end{center}

    \item $\hat{h}_{1;3}=h\sigma^+_{\bm{Q}}+h^*\sigma^-_{\bm{Q}}$
    \begin{center}
    \begin{tabular}{lc}
        \toprule
        index  & operator \\
        \midrule
        \parbox{0.075\textwidth}{cvvv} & \parbox{0.35\textwidth}{$w_{\bold{c}_1\bold{v}_1\bold{v}_1\bold{v}_1}\beta_{\bold{c}_1\bold{v}_1}^\dagger+\rm{h.c.}$}\\
        \parbox{0.075\textwidth}{vccc} & \parbox{0.35\textwidth}{$w_{\bold{v}_1\bold{c}_1\bold{c}_1\bold{c}_1}\beta_{\bold{c}_1\bold{v}_1}+\rm{h.c.}$}\\
        \parbox{0.075\textwidth}{ccvv} & \parbox{0.35\textwidth}{$w_{\bold{c}_1\bold{c}_1\bold{v}_1\bold{v}_1}\beta^\dagger_{\bold{c}_1\bold{v}_1}+\rm{h.c.}$}\\
        \bottomrule
    \end{tabular}
    \end{center}

\end{itemize}

\subsection{two-qubit operator}

\begin{itemize}
    \item $\hat{h}_{2;1}=h\begin{bmatrix}1 & 0 \\ 0 & 0 \end{bmatrix}_{\bm{Q}_1}\otimes \begin{bmatrix}0 & 0 \\ 0 & 1 \end{bmatrix}_{\bm{Q}_2}$
    \begin{center}
    \begin{tabular}{lc}
        \toprule
        index & operator \\
        \midrule
        \parbox{0.075\textwidth}{vvvv\\} & \parbox{0.35\textwidth}{$\frac{n_s}{n_k^2n_c^2}w_{\bold{v}_1\bold{v}_2\bold{v}_1\bold{v}_2}\beta_{\bold{c}_1\bold{v}_1}\beta^\dagger_{\bold{c}_1\bold{v}_1}\beta^\dagger_{\bold{c}_2\bold{v}_2}\beta_{\bold{c}_2\bold{v}_2}$\\$+\frac{n_s}{n_k^2n_c^2}w_{\bold{v}_1\bold{v}_2\bold{v}_2\bold{v}_1}\beta_{\bold{c}_1\bold{v}_1}\beta^\dagger_{\bold{c}_1\bold{v}_2}\beta^\dagger_{\bold{c}_1\bold{v}_1}\beta_{\bold{c}_1\bold{v}_2}$} \\
        \parbox{0.075\textwidth}{cccc\\} & \parbox{0.35\textwidth}{$\frac{n_s}{n_k^2n_v^2}w_{\bold{c}_1\bold{c}_2\bold{c}_1\bold{c}_2}\beta^\dagger_{\bold{c}_1\bold{v}_1}\beta_{\bold{c}_2\bold{v}_2}\beta^\dagger_{\bold{c}_2\bold{v}_2}\beta_{\bold{c}_1\bold{v}_1}$\\$+\frac{n_s}{n_k^2n_v^2}w_{\bold{c}_1\bold{c}_2\bold{c}_2\bold{c}_1}\beta^\dagger_{\bold{c}_1\bold{v}_1}\beta_{\bold{c}_2\bold{v}_1}\beta^\dagger_{\bold{c}_2\bold{v}_1}\beta_{\bold{c}_1\bold{v}_1}$}\\
        \bottomrule
    \end{tabular}
    \end{center}

    \item $\hat{h}_{2;2}=h\sigma^+_{\bm{Q}_1}\otimes \sigma^-_{\bm{Q}_2}+h^*\sigma^-_{\bm{Q}_1}\otimes \sigma^+_{\bm{Q}_2}$
    \begin{center}
    \begin{tabular}{lc}
        \toprule
        index & operator \\
        \midrule
        \parbox{0.075\textwidth}{vvvv\\ \\} &\parbox{0.35\textwidth}{$(\frac{1}{n_kn_cn_s}w_{\rm{mf};\bold{v}_1\bold{v}_2}^v
        +\frac{1}{n_k^2n_c^2}w_{\bold{v}_1\bold{v}_1\bold{v}_2\bold{v}_1}$\\
        $+\frac{1}{n_k^2n_c^2}w_{\bold{v}_1\bold{v}_1\bold{v}_1\bold{v}_2})\beta_{\bold{c}_1\bold{v}_1}\beta^\dagger_{\bold{c}_1\bold{v}_2}+\rm{h.c.}$} \\
        \parbox{0.075\textwidth}{cvcv\\}& \parbox{0.35\textwidth}{$-\frac{1}{n_kn_c}w_{\bold{c}_1\bold{v}_1\bold{c}_2\bold{v}_1}\beta^\dagger_{\bold{c}_1\bold{v}_2}\beta_{\bold{c}_2\bold{v}_2}$\\
        $+w_{\bold{c}_1\bold{v}_1\bold{c}_2\bold{v}_2}\beta^\dagger_{\bold{c}_1\bold{v}_2}\beta_{\bold{c}_2\bold{v}_1}+\rm{h.c.}$}\\
        \parbox{0.075\textwidth}{vccv\\} &\parbox{0.35\textwidth}{$\frac{n_s}{n_kn_c}w_{\bold{v}_1\bold{c}_1\bold{c}_1\bold{v}_2}\beta_{\bold{c}_2\bold{v}_1}\beta^\dagger_{\bold{c}_2\bold{v}_2}$\\
        $-w_{\bold{v}_1\bold{c}_1\bold{c}_2\bold{v}_2}\beta_{\bold{c}_2\bold{v}_1}\beta^\dagger_{\bold{c}_1\bold{v}_2}+\rm{h.c.}$} \\
        \parbox{0.075\textwidth}{cvvc\\} &\parbox{0.35\textwidth}{$\frac{n_s}{n_kn_v}w_{\bold{c}_1\bold{v}_1\bold{v}_1\bold{c}_2}\beta_{\bold{c}_1\bold{v}_2}\beta^\dagger_{\bold{c}_2\bold{v}_2}$\\
        $-w_{\bold{c}_1\bold{v}_2\bold{v}_1\bold{c}_2}\beta_{\bold{c}_1\bold{v}_1}\beta^\dagger_{\bold{c}_2\bold{v}_2}+\rm{h.c.}$} \\
        \parbox{0.075\textwidth}{cccc\\ \\} &\parbox{0.35\textwidth}{$(\frac{1}{n_kn_vn_s}w^c_{\rm{mf};\bold{c}_1\bold{c}_2}
        +\frac{1}{n_k^2n_v^2}w_{\bold{c}_1\bold{c}_1\bold{c}_1\bold{c}_2}$\\
        $+\frac{1}{n_k^2n_v^2}w_{\bold{c}_1\bold{c}_1\bold{c}_2\bold{c}_1})\beta^\dagger_{\bold{c}_1\bold{v}_1}\beta_{\bold{c}_2\bold{v}_1}+\rm{h.c.}$} \\        
        \bottomrule
    \end{tabular}
    \end{center}

    \item $\hat{h}_{2;3}=\begin{bmatrix} 0 & 0 \\ 0 & 1\end{bmatrix}_{\bm{Q}_1}\otimes (h\sigma^+_{\bm{Q}_2}+h^*\sigma^-_{\bm{Q}_2})$
    \begin{center}
    \begin{tabular}{cl}
        \toprule
        index & operator \\
        \midrule
        \parbox{0.075\textwidth}{cvvv\\} & \parbox{0.35\textwidth}{$\frac{n_s}{n_kn_c} w_{\bold{c}_1\bold{v}_2\bold{v}_2\bold{v}_1}\beta_{\bold{c}_2\bold{v}_2}\beta_{\bold{c}_2\bold{v}_2}^\dagger \beta_{\bold{c}_1\bold{v}_1}^\dagger$\\$
        +\frac{1}{n_kn_c}w_{\bold{c}_1\bold{v}_2\bold{v}_1\bold{v}_2}\beta_{\bold{c}_1\bold{v}_2}\beta_{\bold{c}_1\bold{v}_1}^\dagger \beta_{\bold{c}_1\bold{v}_2}^\dagger$}\\
        \bottomrule
    \end{tabular}
    \end{center}
    
    \item $\hat{h}_{2;4}=\begin{bmatrix} 1 & 0 \\ 0 & 0\end{bmatrix}_{\bm{Q}_1}\otimes (h\sigma^+_{\bm{Q}_2}+h^*\sigma^-_{\bm{Q}_2})$
    \begin{center}
    \begin{tabular}{cl}
        \toprule
        index& operator \\
        \midrule
        \parbox{0.075\textwidth}{vccc\\} & \parbox{0.35\textwidth}{$\frac{n_s}{n_kn_v}w_{\bold{v}_1\bold{c}_2\bold{c}_2\bold{c}_1}\beta_{\bold{c}_2\bold{v}_2}^\dagger \beta_{\bold{c}_2\bold{v}_2}\beta_{\bold{c}_1\bold{v}_1}$\\$
        +\frac{1}{n_kn_v}w_{\bold{v}_1\bold{c}_2\bold{c}_1\bold{c}_2}\beta_{\bold{c}_2\bold{v}_1}^\dagger \beta_{\bold{c}_1\bold{v}_1}\beta_{\bold{c}_2\bold{v}_1}$}\\
        \bottomrule
    \end{tabular}
    \end{center}

    \item $\hat{h}_{2;5}=h\sigma^+_{\bm{Q}_1}\otimes \sigma^+_{\bm{Q}_2}+h^*\sigma^-_{\bm{Q}_1}\otimes \sigma^-_{\bm{Q}_2}$
    \begin{center}
    \begin{tabular}{cl}
        \toprule
        index  & operator \\
        \midrule
        \parbox{0.075\textwidth}{ccvv} & \parbox{0.35\textwidth}{$2w_{\bold{c}_1\bold{c}_2\bold{v}_2\bold{v}_1}\beta^\dagger_{\bold{c}_1\bold{v}_1}\beta^\dagger_{\bold{c}_2\bold{v}_2}$}\\
        \bottomrule
    \end{tabular}
    \end{center}

\end{itemize}

\subsection{three-qubit operator}
\begin{itemize}
    \item $\hat{h}_{3;1}=\begin{bmatrix}0 & 0 \\ 0 & 1\end{bmatrix}_{\bm{Q}_1}\otimes(h\sigma^+_{\bm{Q}_2}\otimes\sigma^-_{\bm{Q}_3}+h^*\sigma^+_{\bm{Q}_2}\otimes\sigma^-_{\bm{Q}_3})$
    \begin{center}
    \begin{tabular}{cl}
        \toprule
        index  & operator \\
        \midrule
        \parbox{0.075\textwidth}{vvvv\\} & \parbox{0.35\textwidth}{$\frac{1}{n_k^2n_c^2}w_{\bold{v}_1\bold{v}_2\bold{v}_1\bold{v}_3}\beta_{\bold{c}_1\bold{v}_1}\beta^\dagger_{\bold{c}_1\bold{v}_1}\beta_{\bold{c}_2\bold{v}_2}\beta^\dagger_{\bold{c}_2\bold{v}_3}$\\$
        +\frac{1}{n_k^2n_c^2}w_{\bold{v}_1\bold{v}_2\bold{v}_3\bold{v}_1}\beta_{\bold{c}_1\bold{v}_1}\beta^\dagger_{\bold{c}_1\bold{v}_3}\beta_{\bold{c}_1\bold{v}_2}\beta^\dagger_{\bold{c}_1\bold{v}_1}$}\\
        \parbox{0.075\textwidth}{cccc\\} & \parbox{0.35\textwidth}{$\frac{1}{n_k^2n_v^2}w_{\bold{c}_1\bold{c}_2\bold{c}_2\bold{c}_3}\beta^\dagger_{\bold{c}_1\bold{v}_1}\beta_{\bold{c}_2\bold{v}_1}\beta^\dagger_{\bold{c}_2\bold{v}_1}\beta_{\bold{c}_3\bold{v}_1}$\\$
        +\frac{1}{n_k^2n_v^2}w_{\bold{c}_1\bold{c}_2\bold{c}_3\bold{c}_2}\beta^\dagger_{\bold{c}_1\bold{v}_1}\beta_{\bold{c}_2\bold{v}_2}\beta^\dagger_{\bold{c}_2\bold{v}_2}\beta_{\bold{c}_3\bold{v}_1}$}\\
        \bottomrule
    \end{tabular}
    \end{center}
    \item $\hat{h}_{3;2}=\begin{bmatrix}1 & 0 \\ 0 & 0\end{bmatrix}_{\bm{Q}_1}\otimes(h\sigma^+_{\bm{Q}_2}\otimes\sigma^-_{\bm{Q}_3}+h^*\sigma^+_{\bm{Q}_2}\otimes\sigma^-_{\bm{Q}_3})$
    \begin{center}
    \begin{tabular}{cl}
        \toprule
        index & operator \\
        \midrule
        \parbox{0.075\textwidth}{vvvv\\} & \parbox{0.35\textwidth}{$\frac{1}{n_k^2n_c^2}w_{\bold{v}_1\bold{v}_2\bold{v}_3\bold{v}_2}\beta_{\bold{c}_1\bold{v}_1}\beta^\dagger_{\bold{c}_2\bold{v}_2}\beta_{\bold{c}_2\bold{v}_2}\beta^\dagger_{\bold{c}_1\bold{v}_3}$\\$
        +\frac{1}{n_k^2n_c^2}w_{\bold{v}_1\bold{v}_2\bold{v}_2\bold{v}_3}\beta_{\bold{c}_1\bold{v}_1}\beta^\dagger_{\bold{c}_1\bold{v}_3}\beta^\dagger_{\bold{c}_1\bold{v}_2}\beta_{\bold{c}_1\bold{v}_2}$}\\
        \parbox{0.075\textwidth}{cccc\\ } & \parbox{0.35\textwidth}{$\frac{1}{n_k^2n_v^2}w_{\bold{c}_1\bold{c}_2\bold{c}_1\bold{c}_3}\beta^\dagger_{\bold{c}_1\bold{v}_1}\beta_{\bold{c}_1\bold{v}_1}\beta_{\bold{c}_3\bold{v}_2}\beta^\dagger_{\bold{c}_2\bold{v}_2}$\\$
        +\frac{1}{n_k^2n_v^2}w_{\bold{c}_1\bold{c}_2\bold{c}_3\bold{c}_1}
		\beta^\dagger_{\bold{c}_1\bold{v}_1}\beta_{\bold{c}_1\bold{v}_1}\beta_{\bold{c}_3\bold{v}_1}\beta^\dagger_{\bold{c}_2\bold{v}_1}$}\\
        \bottomrule
    \end{tabular}
    \end{center}
    \item $\hat{h}_{3;3}=h\sigma^+_{\bm{Q}_1}\otimes\sigma^-_{\bm{Q}_2}\otimes\sigma^+_{\bm{Q}_3}+h^*\sigma^-_{\bm{Q}_1}\otimes\sigma^+_{\bm{Q}_2}\otimes\sigma^-_{\bm{Q}_3}$
    \begin{center}
    \begin{tabular}{cl}
        \toprule
        index & operator \\
        \midrule
        \parbox{0.075\textwidth}{cvvv} & \parbox{0.35\textwidth}{$\frac{1}{n_kn_c}w_{\bold{c}_3\bold{v}_1\bold{v}_2\bold{v}_3}\beta_{\bold{c}_2\bold{v}_1}\beta^\dagger_{\bold{c}_2\bold{v}_2}\beta^\dagger_{\bold{c}_3\bold{v}_3}$}\\
        \parbox{0.075\textwidth}{vccc} & \parbox{0.35\textwidth}{$\frac{1}{n_kn_v}w_{\bold{v}_3\bold{c}_1\bold{c}_2\bold{c}_3}\beta^\dagger_{\bold{c}_1\bold{v}_2}\beta_{\bold{c}_2\bold{v}_2} \beta_{\bold{c}_3\bold{v}_3}$}\\
        \bottomrule
    \end{tabular}
    \end{center}
\end{itemize}

\subsection{four-qubit operator}
\begin{itemize}
    \item $\hat{h}_{4;1}=h\sigma^+_{\bm{Q}_1}\otimes\sigma^-_{\bm{Q}_2}\otimes\sigma^+_{\bm{Q}_3}\otimes\sigma^-_{\bm{Q}_4}+h^*\sigma^-_{\bm{Q}_1}\otimes\sigma^+_{\bm{Q}_2}\otimes\sigma^-_{\bm{Q}_3}\otimes\sigma^+_{\bm{Q}_4}$
    \begin{center}
        \begin{tabular}{cl}
            \toprule
            index & operator \\
            \midrule
            \parbox{0.075\textwidth}{vvvv} & \parbox{0.35\textwidth}{$\frac{1}{n_k^2n_c^2}w_{\bold{v}_1\bold{v}_2\bold{v}_3\bold{v}_4}\beta_{\bold{c}_1\bold{v}_1}\beta^\dagger_{\bold{c}_1\bold{v}_3}\beta_{\bold{c}_2\bold{v}_2}\beta^\dagger_{\bold{c}_2\bold{v}_4}$}\\
            \parbox{0.075\textwidth}{cccc} & \parbox{0.35\textwidth}{$\frac{1}{n_k^2n_v^2}w_{\bold{c}_1\bold{c}_2\bold{c}_3\bold{c}_4}\beta^\dagger_{\bold{c}_1\bold{v}_1}\beta_{\bold{c}_3\bold{v}_1}\beta^\dagger_{\bold{c}_2\bold{v}_2}\beta_{\bold{c}_4\bold{v}_2}$}\\
            \bottomrule
        \end{tabular}
    \end{center}
\end{itemize}

\section{Antisymmetrization of the boson}
\label{sec:antisymmetrization}
We introduce a method using the Marumori operator $U_M$ \cite{Marumori_1964} to project the bosonic Hilbert space onto a physical subspace where multiply occupaied states are excluded.    
Suppose that we are considering a two-band system with spin degrees of freedom, $U_M$ is then defined by:
\begin{eqnarray}
    U_M\equiv\ket{0}_f\bra{0}_f\sum_{n=0}C_n(\sum_{\sigma_1,\sigma_2}b^\dagger_{c_{\sigma_1}v_{\sigma_2}}c_{c_{\sigma_1}}c^\dagger_{v_{\sigma_2}})^n\ket{0}_b\bra{0}_b\nonumber \\
    \label{eq:marumori}
\end{eqnarray}
where $C_0=1$, $C_1=1$, $C_2=\frac{1}{2\sqrt{2}}$.
Here $\ket{0}_f$ is the fermion vacuum and $\ket{0}_b$ is the boson vacuum.
$U_M$ transforms a fermionic state into a physical bosonic state as:
\begin{eqnarray}
    &&U_M c^\dagger_{c_{\uparrow}}c^\dagger_{c_{\downarrow}}c_{v_{\uparrow}}c_{v_{\downarrow}}\ket{0}_f\ket{0}_b\nonumber \\
    &&=\ket{0}_f\frac{1}{\sqrt{2}}(
        \beta^\dagger_{c{\uparrow}v{\uparrow}}\beta^\dagger_{c{\downarrow}v{\downarrow}}-\beta^\dagger_{c{\uparrow}v{\downarrow}}\beta^\dagger_{c{\downarrow}v{\uparrow}})\ket{0}_b.
\end{eqnarray}
In this example a fermionic two-electron-two-hole state is transformed into an antisymmetrized two-boson state.
$U_M$ also transforms a fermionic operator into a corresponding bosonic operator as follows:
\begin{eqnarray}
    &&U_Mc_{c\uparrow}^\dagger c_{v\uparrow}U_M^\dagger\nonumber \\
    &&=\{\beta^\dagger_{c{\uparrow}v{\uparrow}}\ket{0}_b\bra{0}_b\nonumber \\
    &&+\frac{1}{\sqrt{2}}(
        \beta^\dagger_{c{\uparrow}v{\uparrow}}\beta^\dagger_{c{\downarrow}v{\downarrow}}-\beta^\dagger_{c{\uparrow}v{\downarrow}}\beta^\dagger_{c{\downarrow}v{\uparrow}})\ket{0}_b\bra{0}_b\beta_{c{\downarrow}v{\downarrow}}\}\ket{0}_f\bra{0}_f.\nonumber \\
\end{eqnarray}
By using $U_M$ we can recover the antisymmetry that was lost through bosonization. 
On the other hand it is clear that the antisymmetrization requires a superposition of all bosonic states.
It spoils our original intention to compress circuit depth.

\bibliography{reference}

\end{document}